\documentclass{aa}
\usepackage{graphicx}
\begin{document}
%%%%%%%%%%%%%%%%%%%%%%%%%%%%%%%%%%%%
%% Macros
\def\wig#1{\mathrel{\hbox{\hbox to 0pt{%
          \lower.6ex\hbox{$\sim$}\hss}\raise.4ex\hbox{$#1$}}}}
\def\rcrit{R_{\rm c}}
\def\cld{{\rm cd}}
%%%%%%%%%%%%%%%%%%%%%%%%%%%%%%%%%%%%%

\title{Evolution of protoplanetary disks:\\
Constraints from DM Tauri and GM Aurigae}

\titlerunning{Protoplanetary disks evolution: DM Tau and GM Aur}
\authorrunning{Hueso and Guillot}

\author{R. Hueso
        \inst{1}
        \and
        T. Guillot
        \inst{2}}

\institute{F\'{\i}sica Aplicada I, E.T.S. Ing.
           Universidad del Pa\'{i}s Vasco, Alda. Urquijo s/n,
           48013, Bilbao, Spain.
           \and
           Laboratoire Cassiop\'ee, CNRS UMR 6202,
       Observatoire de la C\^ote d'Azur, Nice, France.
           }
\offprints{R. Hueso wubhualr@lg.ehu.es or T. Guillot
guillot@obs-nice.fr}

%\date{DRAFT VERSION: \today}
%\date{\today}
\date{Submitted 26 August 2004 / Accepted 21 June 2005}

\abstract{We present a one-dimensional model of the formation and
  viscous evolution of protoplanetary disks. The formation of the
  early disk is modeled as the result of the gravitational collapse of
  an isothermal molecular cloud. The disk's viscous evolution is
  integrated according to two parameterizations of turbulence: The
  classical $\alpha$ representation and a $\beta$ parameterization,
  representative of non-linear turbulence driven by the keplerian
  shear. We apply the model to DM Tau and GM Aur, two classical
  T-Tauri stars with relatively well-characterized disks, retrieving
  the evolution of their surface density with time.
  We perform a systematic Monte-Carlo exploration of the parameter space (i.e.
  values of the $\alpha$-$\beta$ parameters, and of the temperature
  and rotation rate in the molecular cloud) to find the values that
  are compatible with the observed disk surface density distribution,
  star and disk mass, age and present accretion rate. We find that the
  observations for DM Tau require $0.001<\alpha<0.1$ or $2\times
  10^{-5}<\beta<5\times 10^{-4}$. For GM Aur, we find that the
  turbulent viscosity is such that $4\times 10^{-4}<\alpha<0.01$ or
  $2\times 10^{-6}<\beta<8\times 10^{-5}$.
  %, but the additional constraint that the specific initial angular momentum should be
  %$j_{\cld}> 10^{20} \rm cm^2\,s^{-1}$ (as inferred for Class I
  % protostellar envelopes) implies $0.006<\alpha<0.06$ or $\beta\sim
  %10^{-5}$.
  These relatively large values show that an efficient
  turbulent diffusion mechanism is present at distances larger than
  $\sim 10\,$AU. This is to be compared to studies of the variations
  of accretion rates of T-Tauri stars versus age that mostly probe the
  inner disks, but also yield values of $\alpha\sim 0.01$. We show
  that the mechanism responsible for turbulent diffusion at large
  orbital distances most probably cannot be convection because of its
  suppression at low optical depths.

}

\maketitle

\section{Introduction}

The presence of circumstellar disks has been proposed long ago as
a necessary preliminary of star formation, simply because the
decrease of the moment of inertia of a collapsing molecular cloud
core by a factor $10^{16}$ prevents the direct formation of a
central compact object (the star) without a significant loss of
angular momentum. These disks are now routinely discovered, and
both statistical information of disk properties obtained from the
spectral energy distributions and characterization of individual
disks from direct measurements can be obtained (see {\it
Protostars \& Planets IV} for detailed reviews on the subject).

The basic principle yielding the formation of a disk around a
protostar also governs its evolution: material is allowed to be
accreted from the disk onto the protostar only by decreasing its
specific angular momentum. The conservation of mass and total
angular momentum implies that some of the material in the disk
must be transported outward in order for accretion to be possible
(e.g. Lynden-Bell and Pringle, 1974; Pringle, 1981; Cassen, 1994).

One of the major puzzles in understanding circumstellar disks
remains the mechanism by which angular momentum is transported. It
is easy to show that viscosity on a microscopic scale is unable to
explain the dissipation of disks on any reasonable timescale.
Beyond that, the problem resides not in finding a mechanism of
 angular momentum transport but rather on deciding which one is
appropriate and what are the magnitudes of the resulting angular
momentum transport and associated heat dissipation (see e.g.
Cassen 1994; Lin and Papaloizou 1996; Stone et al. 2000). It is
generally believed that disks of relatively small masses transport
their angular momentum by turbulence and that their evolution can
be calculated using an effective turbulent viscosity. The
parameterization of this viscosity then allows the calculation of
the evolution of circumstellar disks to be performed over the
timescales of interest (Myrs).

In this paper, we confront a 1-dimensional theoretical model of
the formation and evolution of circumstellar disks to direct
observations of the disks around the T-Tauri stars DM Tauri and GM
Aurigae. These disks have the particularity that their outer
surface densities (between $\sim 50$ and 800 AU) are constrained
by millimeter-wavelengths observations (Guilloteau and Dutrey
1998; Dutrey et al. 1998), continuum IR emission from the dust
(Kitamura et al. 2002), and %, with much lower precision, by
observations of star light scattered by the disk (Schneider et al.
2003). We use two different parameterizations of the turbulent
viscosity (named $\alpha$ and $\beta$, see \S 3 hereafter) and two
models of molecular cloud collapse.

The simplicity of our model allows for an extensive exploration of
the space of parameters. In total more than 10,000 models were
run. We can therefore constrain the magnitude of the turbulent
viscosity necessary to produce disks of the observed age, mass,
and surface density. The observational constraints however are
subject to very large uncertainties and instead of one simple
scenario of formation and evolution for each disk we find several
kind of solutions to the current state of DM Tau and GM Aur. Our
results can be used to address the questions of the source of the
turbulence, the initial conditions and the type of collapse of the
molecular cloud. They are also helpful in providing a framework
for planet formation, as all the important quantities such as
densities and temperatures can be calculated as a function of time
and distance to the central star.

The structure of this article is as follows: In Section 2 we
present the observational characteristics of the DM Tau and GM Aur
systems, and in particular the parameters that must be fitted by
the models. The numerical model itself and the physics behind it
(including the possible sources of angular momentum transport) are
presented in Section 3. Section 4 presents some model examples and
the strategy employed to fit the observations. We present our
selected sets of models in section 5 and discuss the consequences
for the disks global evolution, the turbulent angular momentum
transport and the initial type of collapse. Finally, section 6
summarizes the main results of the article.

%A second paper will
%present the consequences for planet formation in DM Tau and GM
%Aur.

%%%%%%%%%%%%%%%%%%%%%

\section{Observations of circumstellar disks: DM Tau and GM Aur}
\label{sect:Obs}

DM Tau and GM Aur are both located in the Taurus-Auriga region, a
young stellar association relatively close to Earth, $\sim
140$\,pc (Kenyon et al. 1994), with a low abundance of massive
stars and a relatively cold ($\sim 10-20\,$K) environment (van
Dishoeck et al. 1993). These stars have been chosen for this study
because their disks have been characterized at millimeter to
ultraviolet wavelengths, and because the emission of CO and its
isotomers has been detected by millimetric interferometers to
large distances (500 to 800 AU) to the stars. For these disks, we
thus have access to observations sensitive to the presence of dust
(spectral energy distribution in the infrared and at millimetric
wavelength, visible images of reflected stellar light, images in
the millimetric continuum), to the presence of hot inner gas
accreted by the star (excess emission mainly in the U band) and to
the presence of cold gaseous CO in the outer parts of the disks
(interferometric $^{12}$CO and $^{13}$CO emission maps).

\subsection{DM Tau}

DM~Tau is a low mass ($0.50\rm\,M_{\odot}$) relatively old T-Tauri
star ($\sim 5\times 10^{6}$ yr). Its gas disk was discovered by
Guilloteau and Dutrey (\cite{GD94}) with the IRAM 30-m telescope
and independently by Handa et al. (1995). Further observations and
a $\chi^{2}$ analysis (Guilloteau and Dutrey \cite{Guilloteau98})
allowed the determination of the disk inclination and showed a
rotation rate consistent with a keplerian profile. More recently,
Dartois et al. (2003) combined observations of CO lines of
different C-isotopes to better infer the surface densities and
also estimate the disk vertical temperature profile. Their study
confirmed theoretical expectations that the disk mid-plane at
$\sim 100$\,AU is colder ($T\sim 15$\,K) than 2-3 scale heights
higher ($T\sim 30$\,K).  The disk radius, $850\,$AU, and the slope
of the density profile, $p=1.5$ derived by Guilloteau and Dutrey
differ from a previous study by Saito et al (1995) who derived
$350\,$AU and $p=2.0 \pm 0.3$. The discrepancy can probably be
attributed to the lower resolution (5'' compared to 3'') and
sensitivity of the observations by Saito et al.

The outer disk radius measured for DM Tau either from the mm
continuum emission (Kitamura et al. \cite{Kitamura}) or from HST
near IR coronographic images (Grady et al. \cite{Grady}) is
assumed to be a lower limit because of the systematically smaller
lower sensitivities (in terms of equivalent surface densities) of
these techniques compared to the CO observations.

Recent observations of DM Tau in the near- to mid-infrared
($2.9-13.5\,\mu$m) and preliminary radiative transfer models by
Bergin et al. (2004) seem to indicate that the inner region of the
disk consists of three parts: (i) close to the star, and up to
$\sim 4$\,AU, a flat, optically thin disk, cleared of most of its
dust (and probably gas); (ii) At 4\,AU, a ``wall'', heated by
direct stellar irradiation to $\sim 120$\,K, and extending
vertically to $\pm 0.5$\,AU relative to the midplane; (iii) A
``standard'' optically thick disk beyond that distance.

\subsection{GM Aurigae}

The evidence of the presence of a disk around GM Aur was first
obtained from radiometric observations by Koerner et al. (1993).
Higher resolution interferometric maps were later obtained by
Dutrey et al. (1998), and Kitamura et al. (2002) using the same
techniques as for DM Tau. Depending on the inclination of
the disk, two solutions can be found for the star mass, i.e.
$M_{*}=0.5$ or 0.9$\,\rm M_\odot$ with most probable values close
to 0.9. Its disk appears to be slightly smaller than that of DM
Tau, and have an outer radius $\sim 525$\,AU, as determined from
$^{12}$CO observations. High-resolution coronographic HST images
of scattered near-infrared light (Wood et al. 2002; Schneider et
al. 2003) indicate that the disk is outwardly flared and extends
at least to 300\,AU.

As for DM Tau, the near- to mid-infrared observations indicate the
presence of an inner disk clearing within ~6.5\,AU, with a 120\,K
``wall'' at that distance and an optically thick disk beyond (Bergin
et al. 2004). The presence of such a gap (but a slightly smaller, i.e.
$\sim 4$\,AU one) was also an outcome of the modeling of a less
accurate SED (Chiang and Goldreich 1999, Wood et al. 2002), and
attributed to the presence of a $\sim 2\rm\,M_{J}$ planet orbiting at
2.5 AU (Wood et al, 2002; Rice et al. 2003). This explanation remains
mostly speculative at this point, however.

\subsection{Inferred constraints}
\label{sec:constraints}

\begin{table}[tb]
         \begin{tabular}{p{2.7cm} p{2.4cm} p{2.4cm}lcc}
            \hline
            \vspace{0.005in}\textbf{ System:} \smallskip        & \vspace{0.005in} \textbf{DM Tau} \smallskip         & \vspace{0.005in} \textbf{GM Aur}  \smallskip         \\
            \hline
            Distance:       & 140 pc $\pm$ 10 \%    & 140 pc $\pm$ 10 \%    \\
            Age:            & $1.5-7$       Myr     & $1-10$           Myr  \\
            Spectral Type:  & M1                    & K7                    \\
            Star Mass:      & 0.4-0.6 M$_{\odot}$   & 0.5-1.0 M$_{\odot}$   \\
            Luminosity:     & 0.25 L$_{\odot}$      & 0.74 L$_{\odot}$      \\
            Temperature:    & 3720 K                & 4060 K                \\
            Star Radius:    & 0.0065 AU             & 0.0085 AU             \\
            \hline
        \end{tabular}
        \begin{tabular}{p{7.8cm} p{1.4cm} lc}
        \vspace{1mm} \textbf{Constraints from CO emission lines:} & \\
        \end{tabular}
        \begin{tabular}{p{2.0 cm} p{0.3cm} p{2.4cm} p{2.4cm} lccc}
           $R_{out}$ & $\sim$ & $850\pm 20$ AU  &  $525\pm 20$ AU  \\
           $\Sigma_{(100 AU)}$         & $>$    & 0.05\,g\,cm$^{-2}$ & $-$  \\
           $\Sigma_{(R_{out})}$        & $>$    & $1.6 \times 10^{-3}$\,g\,cm$^{-2}$ & $3.0 \times 10^{-3}$\,g\,cm$^{-2}$  \\
        \end{tabular}
        \begin{tabular}{p{7.8cm} p{1.4cm} lc}
        \vspace{1mm} \textbf{Constraints from continuum dust emission:} & \\
        \end{tabular}
        \begin{tabular}{p{2.0 cm} p{0.3cm} p{2.4cm} p{2.4cm} lccc}
           $R_{out}$            & $>$ & 500 AU              & 280 AU \\
           $\Sigma_{(100 AU)}$  & $<$ & 4.3  g\,cm$^{-2}$     & 23.0 g\,cm$^{-2}$  \\
           $\Sigma_{(100 AU)}$  & $>$ & 0.23  g\,cm$^{-2}$    & 1.4
           g\,cm$^{-2}$  \\
       $\tilde{\beta}$ && $0.48-1.03$ & $1.11-1.34$  \\
           \hline
        \end{tabular}
        \begin{tabular}{p{5.8cm} p{1.4cm} lc}
        \vspace{1mm} \textbf{Accretion rate:} & \\
        \end{tabular}
        \begin{tabular}{p{2.0 cm} p{0.3cm} p{2.4cm} p{2.4cm} lccc}
           Log $\dot{M}$ & $=$ &
           $-7.95 \pm 0.54$ & $-8.02 \pm 0.54$ \\
        $\quad \rm [M_\odot\,yr^{-1}]$ & & & \\
           \hline
        \end{tabular}
\vspace{1mm} \caption[]{Physical parameters for DM Tau and GM Aur:
Stellar parameters are from Simon et al (2001), CO constraints from
Dartois et al. (2003) and Dutrey et al (1998), dust emission
constraints from Kitamura et al. (2002) (with an additional 1/3 to 3
uncertainty factor --see text) and accretion rates from Hartmann et
al. (1998) and Gullbring et al. (1998). }
\end{table}

We detail hereafter the observational constraints that are directly
relevant to the disk evolution calculations and that are summarized in
Table~1. We choose to adopt relatively conservative constraints. (More
restrictive constraints may be obtained from SED fitting and direct
analysis of the emission maps which is beyond the scope of the present
article.)

\begin{description}
\item[{\it Stellar masses:}] They are inferred from the
   keplerian rotation rates measured in $^{12}$CO (Simon et
   al. \cite{Simon01}). The uncertainty on this parameter is
   essentially due to that of the distance given by Hipparcos to an
   accuracy $\sim 10$\,\%. The inferred stellar masses depend also on the
   the disk inclination.

\item[{\it Ages:}] This is a crucial, but unfortunately very uncertain
  parameter entering the models. The uncertainty on the stellar masses
  propagates into that on the inferred age. Taking that into account,
  and using evolution tracks from Baraffe et al. (1998, 2002), we
  derive for both stars a range of ages that covers ages published by
  different groups (e.g. Guilloteau \& Dutrey 1998; Hartmann et
  al. 1998).

\item[{\it Accretion rates:}] As discussed by Hartmann et al. (1998),
   an accretion luminosity $L_{\rm acc}$ can be derived from the
   measured excess emission observed in the U band. It is then turned
   into an accretion rate following the relation:
\[\dot{M}={1\over
   \gamma}{R_\star\over GM_\star}L_{\rm acc}, \]
   where $R_\star$ and
   $M_\star$ are the radius and mass of the star, respectively, and
   $\gamma$ is a factor that accounts for the distance from which the
   material free-falls onto the star. This parameter is expected to be
   $\sim 0.8$ due to the presence of a magnetic cavity inside
   $5\,R_\star$. The uncertainties on $\dot{M}$ stem both from those
   linked to the determination of the bolometric accretion luminosity,
   and those related to the uncertain $\gamma$. Hartmann et
   al. estimate the uncertainty on $\dot{M}$ to be within a factor 3,
   with more optimistic error estimates leading to a factor $\sim 2$,
   given the likely presence of the magnetic cavity.
\item[{\it Surface densities from measured CO emission maps:}] In
the
   case of DM Tau, measurements of $^{12}$CO, $^{13}$CO and C$^{18}$O
   lines allow for a direct determination of the density profile of gaseous
   CO from orbital distances $\sim 100$\,AU to the outer disk (Dartois
   et al. 2003). Interestingly, the outer radius of the disk is
   different depending on the molecular species that is considered,
   which is interpreted as selective photodissociation in the
   outer disk. $^{12}$CO being the most abundant isotomer, it is also
   the most resistant to photodissociation and yields the largest
   radius (used for this work). Because CO furthermore tends to
   condense in these cold regions (e.g. Aikawa et al. 1999), the gas mass
   can only be thought as a lower limit to the outer disk mass.
   A comparison of CO and dust emission in DM Tau indeed
   yields a disk mass that is smaller
   by a factor $\sim 5$ when inferred from the CO measurement than when
   inferred from the dust (Dartois et al. 2003). This appears to be
   consistent with CO condensation, but may also be partially explained
   by a larger continuum opacity. The same conclusions hold for GM
   Aur, but the set of observations is more limited, because it is
   based on $^{13}$CO $J=1\rightarrow 0$ (Dutrey et al. 1996) and
   $^{12}$CO $J=2\rightarrow 1$ (Dutrey et al. 1998) emission. In
   Table~1, the minimum values of the surface density required to
   explain the CO emission are derived from Dartois et al. (2003) for
   DM Tau and Dutrey et al. (1998) for GM Aur. Conservatively, we
   assumed that all CO was in gaseous form and with a solar abundance
   to derive these minimum densities.
\item[{\it Surface densities from continuum emission maps:}]
Because
   continuum millimetric and sub-millimetric emissions are optically
   thin, their measurements inform us on the surface density and global
   masses of the disks. However, for a given flux, the surface density
   (or equivalently disk mass) derived is inversely proportionnal to an
   uncertain opacity coefficient $\kappa_\nu$. Following Beckwith et
   al. (1990), this opacity is often approximated by:
\begin{equation}
  \kappa_\nu= \kappa_0 \left({\nu\over 10^{12}\
   \mbox{Hz}}\right)^{\tilde{\beta}}, \label{eq:kappa}
\end{equation}
   where $\kappa_0\approx 0.1\rm cm^2\,g^{-1}$ and $\tilde{\beta}$ are
   parameters that depend on the grain composition and size
   (e.g. Beckwith et al. 2000). $\kappa_0$ is the opacity per total
   mass (gas+dust) and therefore depends on an assumed dust to gas
   ratio. The $\tilde{\beta}$ parameter can be obtained
   from the slope of the spectral energy distribution in the mm region
   (e.g. Beckwith et al. 1990; Chiang \& Goldreich et al. 1997).
   Uncertainties on $\kappa_0$ are often neglected in the calculation
   of disk masses and surface densities although they are probably large
   (Beckwith et al. 1990; Pollack et al. 1994; Hartmann et al. 1998;
   D'Alessio et al. 2001).

   In this work, we adopt the matches to the SEDs and
   2 mm emission maps obtained by Kitamura et
   al. (\cite{Kitamura}) to constrain the surface density at 100
   AU. We choose not to use their constraints on the disk masses and
   radii and on the slope of the surface density profile because the
   loss of sensitivity of the observations at large orbital distances
   and the limited spatial resolution imply that these are very
   model-dependant. Kitamura et al. do include a range of allowed
   $\tilde{\beta}$ values to estimate the characteristics of the disks, but do
   not allow for variations in $\kappa_0$. In pratice, the values of
   $\tilde{\beta}$ for DM Tau are smaller (by $\sim 0.45$) than those for GM
   Aur. Given the standard opacity law (eq.~\ref{eq:kappa}), this
   implies that the assumed 1.3\,mm opacity is about twice larger for
   DM Tau than for GM Aur. On the other hand, models of grain growth
   indicate that low values of $\tilde{\beta}\wig{<}1$ are consistent with the
   presence of large grains, and also generally lead to smaller
   opacities in the millimetric (Miyake \& Nakagawa 1993; D'Alessio et al. 2001). This
   implies that the larger disk masses derived for GM Aur may be (at
   least partly) an artefact of the opacity law.

   In order to account for the uncertainty on $\kappa_0$ we hence
   consider cases for which the 100\,AU gas surface density is either
   multiplied or divided by a factor 3 compared to the values
   obtained by Kitamura et al. (2002).
\end{description}

\section{Model description}
Our aim in this paper is to obtain evolutionary models able to
satisfy the observational constraints provided by Table 1 and
perform a sensitivity analysis of the model parameters for DM Tau
and GM Aur. This requires solving the evolution of the gas disk
for $\sim$10 Myrs and searching not only for the "best" models but
for the ensemble of parameters compatible with the observations
and their uncertainties. With current (and near-future) computing
facilities we must restrict ourselves to a relatively simple
system of 1D radial equations in which all quantities have been
vertically averaged and radiative transport is approximated.
Consequences of this averaging appear to be relatively mild in
terms of accuracy (Hur\'e and Galliano, 2001; see also Ruden and
Lin, 1986). This is certainly not a concerning problem, given that
we are seeking order-of-magnitude constraints on the turbulent
viscosity and other relevant parameters.

We hereafter describe our system of equations in the framework of
a simple, but relatively complete model including -hopefully- most
relevant physical mechanisms. We also discuss its limitations in
Section~3.5.

\subsection{Viscous evolution of the surface density}

The evolution of a disk follows conservation of angular momentum
and mass. As mass is accreted into the inner region of the disk
and finally onto the central star, part of the disk material must
migrate outwards to conserve angular momentum. If the disk is
axisymmetric and geometrically thin, the dynamics are only
radially dependent and an effective turbulent viscosity, $\nu$,
can be postulated as the angular momentum transport mechanism. In
this case, the equation giving the evolution of the disk surface
density, $\Sigma$, can be written as,
\begin{equation}
\label{Sigma}
\frac{\partial \Sigma}{\partial t} = \frac{3}{r}
\frac{\partial}{\partial r} \left(\sqrt{r}\frac{\partial}{\partial
r} \left[ \nu \Sigma \sqrt{r} \right] \right)+S(r,t),
\end{equation}
where we have introduced $S(r,t)$, a source term accounting for
the accretion from the molecular cloud core onto the circumstellar
disk (see \S\ref{collapse}).

The general solution of (\ref{Sigma}) is a disk that progressively
accretes mass inwards while redistributing a part of the outermost
material further away to conserve angular momentum until all of the
material is accreted and all of the angular momentum has been
transported away (Lynden-Bell and Pringle, \cite{LBP}). This
process is modulated by the value of the (turbulent) viscosity $\nu$.

\subsection{Two parameterizations of turbulence: $\alpha$
  and $\beta$ models}

A prescription for the turbulent viscosity was originally
postulated by Shakura and Sunyaev (\cite{Shakura73}) and has been
extensively used in models of turbulent disks known since then as
alpha disk models.
According to this prescription,
\begin{equation}
\nu=\alpha c_{s} H.
\label{eq:nu_alpha}
\end{equation}
The free parameter $\alpha$ controls the amount of turbulence in a
turbulent medium where the scale height $H$, and the isothermal
sound speed $c_{s}$, are upper limits to the mixing length and
turbulent velocity, respectively. Accretion rates inferred in T
Tauri stars are compatible with $\alpha\approx 10^{-2}$ (Hartmann
et al, 1998).  As a caveat, one should note that it is difficult
to imagine any physical process that would yield a viscosity
obeying strictly to this parameterization, with $\alpha$ being
spatially uniform and temporally constant. In this sense, any
value of $\alpha$ that can be retrieved has to be considered as an
undefined average over both time and space of the adimensional
quantity $\nu/(c_{s}H)$.

A possible source of angular momentum transport that can be
approximately parameterized in the $\alpha$-framework is due to
the so-called magnetorotational instability (Balbus and Hawley
1991, 1998, 2000). The instability arises from the fact that
ionized elements in a magnetized environment tend to conserve
their velocity. In a keplerian magnetized disk, an element being
displaced inward will have a lower velocity than other elements at
the same location. Therefore, it will also have a smaller angular
momentum and it will tend to sink further in (see Terquem 2002 for
a nice description of the process). The result is an inner
transport of material and an outward transport of angular
momentum. Numerical calculations of rotating magnetic disks
indicate that a magnetorotational instability can effectively
yield a viscosity with $\alpha$ between $10^{-3}$ and $0.1$
(Brandenburg et al.  1999; Stone et al. 2000; Papaloizou \& Nelson
\cite{PN03}).  The suppression of the instability in neutral
regions of the disk could imply rapid spatial variations of
$\alpha$ (Gammie, 1996).

% TG deleted 2 next sentences
%% However, work by Fromang et al.  (2002)
%% suggests that if $\alpha\wig{>}10^{-2}$, the heat dissipated by
%% turbulent viscosity yields an ionization that remains sufficient for
%% instability to occur, justifying the use of a uniform $\alpha$. An
%% unsolved question in our study is the pertinence of the
%% magnetorotational instability in the outer regions of the disk.

Another mechanism initially proposed to be responsible for angular
momentum transport is thermal convection (Lin and Papaloizou,
1980; Ruden and Lin, 1986). The mechanism had largely fallen out
of favor after direct hydrodynamical simulations of thermal
convection in disks by Stone and Balbus (1996; see also Stone et
al. 2000) had resulted in very small angular momentum transport,
mostly inward instead of outward. However, these results are
challenged by recent simulations of convection in a baroclinically
unstable disk that yield an outward angular momentum transport and
corresponding values of $\alpha$ between $10^{-4}$ and $10^{-2}$
(Klahr et al, 1999; Klahr and Bodenheimer, 2001). Turbulent
angular momentum transport due solely to thermal convection can be
tested by our models because of the particularity that it will
cease in an optically thin medium. As proposed by Ruden and
Pollack (1991), turbulent viscosity can then be modeled by a value
of $\alpha$ which is uniform in the optically thick part, and goes
to zero when the disk becomes optically thin.

Another potential source of angular momentum transport is that due
to shear instabilities (e.g. Dubrulle 1993). This mechanism was
claimed to be inadequate on the basis of numerical simulations
(Balbus et al. 1996; Hawley et al. 1999). However,
the disappearance of turbulence in the hydrodynamical simulations
may be an artifact due to the limited resolution of the
simulations (Longaretti, 2002). Indeed, turbulence has been shown
to be sustained in Couette-Taylor experiments with outward
decreasing angular velocity (Richard and Zahn, 1999). A different
prescription for the turbulent velocity may then be applied:
\begin{equation}
\nu=\beta \frac{\partial \Omega}{\partial R} R^{3},
\end{equation}
where $\Omega$ is the disk rotation rate. Richard \& Zahn found
$\beta\sim 2\times 10^{-5}$. Hur\'e et al. (2001) further showed
that disk models using this prescription and $\beta~\sim 10^{-5}$
are as capable of explaining the observed accretion rates and
disks lifetimes as $\alpha$ models.  A convenient feature of this
turbulent parameterization is that $\nu$ does not depend on the
temperature: the disk's evolution then becomes independent of
complex issues related to radiative transfer and opacities.

Other mechanisms can yield angular momentum transport in
circumstellar disks without necessarily being amendable to the
calculation of a turbulent viscosity. This is for example the case
of gravitational instabilities and density waves (e.g. Laughlin and
R\'o$\dot{\mathrm{z}}$yckza, 1996; Laughlin et al. 1998), which may
be an important source of disk evolution early on, when the
disk/star ratio is still relatively large. It is also the case of
bipolar outflows (e.g. Konigl, 1989; Casse and Ferreira, 2000).
These probably also have an important role during the cloud collapse
phase. Although the ejection is limited to $\sim 10$\% of the
material accreting onto the central star (Calvet, 1997), this could
nevertheless represent a substantial sink of angular momentum so
that Eq.~(2) would not be valid in the central region affected by
the outflow. Since both of these effects are of importance only
during the early evolution phases, neglecting them yields very
limited quantitative changes to our results (see \S 5.1).

\subsection{Gravitational collapse of an isothermal molecular cloud
  core and disk growth}
\label{collapse}

%TG: All this section rewritten -many things removed
The formation and early evolution of the star+disk system is governed
by the collapse of its parent molecular cloud core. This process
remains poorly known (e.g. Andr\'e et al. 2000), and we choose to
model it using several simplifying assumptions. First, we assume that
the cloud envelope is isothermal and spherically symmetric. Any given
shell of radius ${\ell}$ and angular velocity $\omega({\ell})$
will collapse onto the disk within the centrifugal radius (the point
at which the maximal angular momentum in the shell is equal to the
angular momentum in the disk):
\begin{equation}
\rcrit(t)=\frac{{\ell}(t)^{4}\omega(\ell)^{2}}{GM(t)},
\label{Rd}
\end{equation}
where $M(t)$ is the mass that has been accreted onto the star+disk
system at time $t$. (In the formalism of the self-similar solutions,
$t=0$ corresponds to the formation of the central
condensation). Formally, eq.~\ref{Rd} is valid only in the limit when
the disk's gravitational potential can be neglected. However, given
our lack of knowledge of the collapse phase, this simplification is
perfectly justified.

We further assume that angular momentum is conserved during the
collapse and that material falling onto the disk finds its way to a
location where its angular momentum is equal to that of the
(keplerian) disk. With that hypothesis, and given $\dot{M}$, the
accretion rate onto the disk, we derive the following source term:
\begin{equation}
S(r,t) = \frac{\dot{M}}{\pi \rcrit^2}\frac{1}{8}
\left(\frac{r}{\rcrit}\right)^{-3/2}
\left[1-\left(\frac{r}{\rcrit}\right)^{1/2}\right]^{-1/2}.
\label{source.guillot}
\end{equation}
Departures from a central gravitational potential during the collapse
phase imply that Eq.~(\ref{Sigma}) does not perfectly conserve the
angular momentum of the disk+envelope system. However, once again, the
effect is negligeable.

\begin{figure}
\centering
\includegraphics[width=8.5cm]{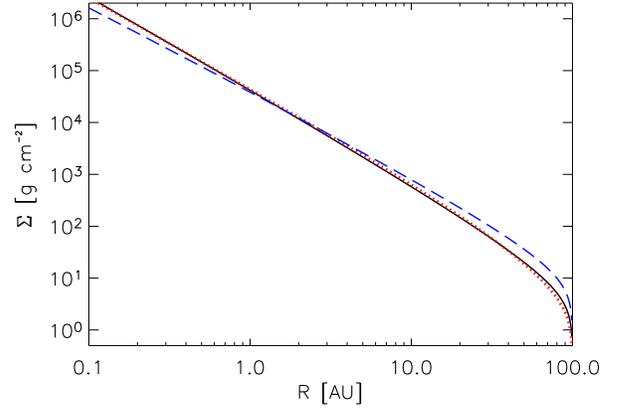}
\caption{Surface density obtained at the end of the collapse of a
  $0.3\,\rm M_0$ disk, assuming a centrifugal radius $\rcrit=100\,$AU,
  and no viscous stress. The plain line corresponds to the collapse of
  a molecular cloud core in solid rotation. The dotted line shows the
  same solution when accounting for the outward angular momentum
  transport due to the subkeplerian momentum of the incoming envelope
  material (see Cassen \& Moosman \cite{Cassen81}). The dashed line
  corresponds to the collapse of magnetized cloud core with a rotation
  rate $\omega(\ell)\propto 1/\ell$.}
\label{Fig.collapse}
\end{figure}

This expression of $S(r,t)$ differs from that obtained from ballistic
integrations by Cassen \& Moosman (1981) who showed that the envelope
material that encounters the disk has in fact a subkeplerian rotation
rate (see also Nakamoto \& Nakagawa \cite{Nakamoto94}). This yields a
small additional, outward, angular momentum transport that we choose
to ignore in this simulation: as shown in
Fig.~\ref{Fig.collapse}\ below, the effect would tend to decrease the
surface density in the outer regions by 40\% at most. The turbulent
viscosities that are considered have a much more significant
effect. Furthermore, this is also relatively small compared to most
sources of uncertainties, in particular those related to the cloud
collapse itself.

The values of $\dot{M}$ and $\rcrit(t)$ depend on the mechanisms
that led to the collapse of the molecular cloud. A simple and widely
used solution is that of Shu (1977). In that case, a self-similar
approach to the problem with $\omega(\ell)=\omega_{cd}$, a constant
rotational speed of the molecular cloud, shows that the collapse may
proceed from inside-out, with a constant mass accretion:
\begin{equation}
  \dot{M}=0.975 \frac{c_s^{3}}{G},
  \label{eq:mdot}
\end{equation}
where $c_s$ is the isothermal sound speed. Because the collapse
solution yields $\ell=c_s t/2$ and using a mean molecular weight
$\mu=2.2$, one can show that the centrifugal radius can be
written:
\begin{equation}
\rcrit(t) \simeq 53
\left(\frac{\omega_{\cld}}{10^{-14}\mathrm{s^{-1}}}\right)^2
\left(\frac{T_{\cld}}{10 \mathrm{ K}} \right)^{-4}
\left(\frac{M(t)}{1 \rm M_{\odot}} \right)^3 \mathrm{AU},
\label{EqRdcanonic}
\end{equation}

 Observations of hot cores indicate that the accretion rate is
not constant over time but enhanced in the very first stages of
cloud collapse and progressively diminishing after a more or less
long time span of accretion rate close to the one predicted by the
Shu theory (Bontemps et al. 1996). However, observations of some
very young protostars like IRAM 04191 in the Taurus molecular cloud
(Belloche 2002) point to the presence of differential rotation and
hence different initial conditions. Indeed, in the presence of a
substantial magnetic field, the collapse phase can begin in a cloud
whose rotation rate is $\omega(\ell)\propto 1/\ell$. In that case,
accretion proceeds much faster, and a self-similar approach shows
that the centrifugal radius then evolves linearly with the accreted
mass (Basu 1998):
\begin{equation}
\dot{M} \simeq 10 \frac{c_{s}^3}{G},
\label{eq:mdot-basu}
\end{equation}

\begin{equation}
\rcrit(t) \simeq 15
\left(\frac{\omega_{b}}{10^{-14}\mathrm{s^{-1}}}\right)^2
\left(\frac{B_{ref}}{30 \mu \mathrm{G}} \right)^{-2} \left(\frac{M(t)}{1
\rm M_{\odot}} \right) \mathrm{AU},
\label{eq:rd-basu}
\end{equation}
where $\omega_{b}$ is the ambient rotation rate of the cloud and
$B_{\rm ref}$ is a background reference magnetic field. The
numerical factor in front of the accretion rate is one possible
value among many but all models imply it is of order 10 (e.g.
Foster \& Chevalier \cite{Foster}, Basu 1998, Hennebelle et al.
2003). Because the accretion rate also depends on the poorly-known
$c_s^3$, this uncertainty is ignored.

In the absence of a viscous stress, it can be shown that the collapse
of the cloud core yields a profile density $\Sigma(r)=\int S(r,t)dt$
that is approximately proportional to $r^{-7/4}$ in the case of a
cloud core in solid rotation and to $r^{-3/2}$ in the case of our
magnetized, differentially rotating cloud core. The resulting profiles
are compared in Fig.~\ref{Fig.collapse}. Because the centrifugal disk
radius at the end of the collapse is fixed in this comparison, the
magnetized cloud core has initially more angular momentum: the disk
that forms therefore has more mass in its outer regions.

\subsection{Calculating disk properties}

Solving Eq.~\ref{Sigma} necessitates the calculation of quantities
such as the mid-plane temperature, $T_{c}$, the vertical scale height,
$H$, and the keplerian rotational frequency $\Omega_{k}$.  This has
been carried out in many papers (see e.g. Ruden and Lin,
\cite{Ruden86}; Ruden and Pollack, \cite{RudenPollack}; Reyes-Ruiz and
Stepinski, \cite{ReyesRuiz} to cite only a few), but the detailed
expressions may differ slightly from one work to another.  Generally,
these works have considered disks of small masses and
sizes, and have neglected the gravitational potential of the disks
themselves. Because this effect may be important for the evolution of
DM Tau and GM Aur, we rederive here vertically averaged equations
including autogravitation.

\subsubsection{Autogravitation and general disk properties}

We follow Ruden and Pollack (\cite{RudenPollack}) in expressing
the isothermal sound speed, $c_{s}$, and the mid-plane density,
$\rho_{c}$ as
\begin{equation}
c^{2}_{s}=\frac{\mathcal{R}}{\mu}T_{c},
\end{equation}
\begin{equation}
\label{Eq_rho}
\rho_{c}=\frac{\Sigma}{2H}.
\end{equation}
Here $\mathcal{R}$ is the gas constant, $\mu$ the molecular weight
of the gas, supposed to be 2.2, and $H$ is a measure of the
vertical scale height, which is a well defined parameter only in
non autogravitating disks. The mass of the disk also affects the
keplerian rotational frequencies of the disk:
\begin{equation}
\label{Eq_Omega} \Omega_{k}(r,t)=\left [ \frac{GM_{*}(t)}{r^3}  +
\frac{1}{r} \frac{dV_{d}}{dr} \right ]^{1/2},
\end{equation}
where $V_{d}$ is the gravitational potential of the disk. This
potential can be calculated by
\begin{equation}
\label{Vd}
V_{d}(r)=\int_{R_{*}}^{\infty} \int_{0}^{2\pi}\frac{-G
\Sigma(r_{1})r_{1}}{\sqrt{(r_{1} \sin \theta)^2+ (r-r_{1}\cos
\theta)^2}}dr_{1}d\theta.
\end{equation}

It is convenient to express Eq. (\ref{Vd}) in terms of the radial
coordinate alone by using the elliptic integral of the first kind
$K[m]$:
\begin{equation}
V_{d}(r)=\int_{R_{*}}^\infty -G
\frac{4\mathrm{K}\left[-\frac{4(1+r/r_{1}-1))}{(r/r_{1}-1)^2}\right]}{
|(r/r_{1}-1)|}\Sigma(r_{1})dr_{1}.
\end{equation}

Generally, the incorporation of $V_{d}$ does not play a major
role in the evolution of the disk. Its effect is to slightly
increase $\Omega_{k}$ decreasing the turbulent viscosities through
the definitions of $H$ and $\nu$. Its influence is greater in
massive and extended disks with an inner low mass star.

The vertical scale height, $H$, is calculated assuming vertical
hydrostatic equilibrium, considering the vertical component of
star gravity and the local gravitation of the disk. This is
approximated from the infinite and R-homogeneous slab
approximation, which considers the locally equivalent limits
$\Sigma$=cte or $z\rightarrow 0$ (Paczy\'nski, 1978; Hur\'e,
2000):
\begin{equation}
\frac{1}{\rho}\frac{dP}{dz}=-\Omega_k^{2}z-4\pi G \Sigma \equiv
gz,
\end{equation}
where $g$ is the effective gravity from both star and disk.

An important simplification is to approximate the disk's vertical
structure by an isotherm (the departures from an isotherm yield only
second order effect). In that case, the disk pressure and density
decrease exponentially:
\begin{equation}
(P,\rho)=(P_{0},\rho_0) \exp \left( -(z/H_{0}+(z/H_{1})^2) \right),
\end{equation}
where the $H_{0}$ term comes from the autogravitation term and
$H_{1}$ is the classical scale height coming from the vertical
component of the radial gravity of the star. They are defined as
\begin{eqnarray}
H_{0}&=&\frac{c^2_{s}}{4 \pi G \Sigma},\\
H_{1}&=&\frac{\sqrt{2}c_{s}}{\Omega}.
\end{eqnarray}

The first one is dominant for large distances and large values of
$\Sigma$, where the disk gravitation dominates over the vertical
component of the central star gravitation. The second one is the
scale height in absence of autogravitation. Consistency with our
definition of $\rho_{c}$ through Eq. (\ref{Eq_rho}) requires that
$H$ has the following form:
\begin{equation}
H=H_{1}\frac{\sqrt{\pi}}{2}\exp \left(\frac{H_{1}}{2H_{0}}
\right)^2 \left( 1-{\rm Erf} \left( \frac{H_{1}}{2H_{0}}\right)
\right)
\end{equation}
When autogravitation is negligible, $H_0$ is large and $H
\rightarrow \sqrt{\pi}/2 H_1$. When the disk's gravity is
significant, $H_0$ is small and $H < H_1$. The role of the disk
gravity is mainly restricted to produce a slight vertical
flattening of the outer disk.

The assumption of vertical isothermal structure is generally accurate
only close to the disk mid-plane and values of $H$ should be
considered only as reasonable estimates instead of exact calculations.
This factor yields a slight uncertainty on the magnitude of $\nu$ in
the case of the $\alpha$ parameterization.

\subsubsection{Temperature calculation}

In order to calculate mid-plane temperatures the following
assumptions are made:
\begin{itemize}
\item The disk is geometrically thin;
\item It is heated by the star's illumination but also by the
  dissipation of viscous energy by turbulence;
\item It is assumed to be optically thick in the radial direction everywhere so that
heat can be transported efficiently only in the vertical direction.
\end{itemize}

In thermal equilibrium, the disk's cooling flux is equal to the sum
of the heating fluxes due to viscous dissipation and external
sources: We define $T_{l}$ as the effective temperature at which the
disk is being heated by the central star and external sources and
$T_e$ as the emission temperature at which the disk cools down.
These quantities are then related by the following expression:
\begin{equation}
2\sigma_{B}T_{e}^{4}=\Sigma \nu r^2 \Omega_{r}^2 +
2\sigma_{B}T_{l}^{4},
\end{equation}
where $\sigma_{B}$ is Stephan-Boltzmann's constant, $T_{e}$ is the
effective temperature of the disk, and $T_{l}$ is an effective
temperature to which an inert disk would be heated by external
sources. $T_{l}$ will be discussed in the next section.

Obtaining the mid-plane temperature $T_{m}$ from the effective
temperature $T_{e}$ is generally a complex radiative transfer
problem. However, in both the optically-thick and optically-thin
regimes, analytic expressions can be found (assuming that the
opacities behave correctly). In the present work, we adopt
the expression derived by Nakamoto and Nakagawa (\cite{Nakamoto94}),
who approximate the mid-plane temperature as a sum of optically-thick
and optically-thin contributions:
\begin{equation}
\label{TEq}
\sigma_{B}T_{c}^{4}=\frac{1}{2}\left(\frac{3}{8}\tau_{R}+
\frac{1}{2\tau_{p}}\right)\Sigma \nu r^2 \Omega_{r}^2 +
\sigma_{B}T_{l}^{4},
\end{equation}
where $\tau_{R}$ and $\tau_{P}$ are the Rosseland and Planck mean
optical depths respectively. The optical depth from the midplane
to the surface of the nebula, $\tau_{R}$, is defined in terms of
the Rosseland mean opacity $\kappa_{R}(\rho_{c},T_{c})$ by
\begin{equation}
\tau_{R}=\frac{\kappa_{R}\Sigma}{2}.
\end{equation}

To evaluate $\kappa_R$, we used the Rosseland mean opacity of dust
grains provided by Pollack et al (1986) in the form of a power
law: $\kappa_{R}(\rho_{c},T_{c})=K_{0}{\rho_{c}}^{m}{T_{c}}^{n}$.
Note that the rapid changes of the opacities in different regimes
(e.g. due to the evaporation of water,...etc.) imply that a robust
autoconsistent numerical method must be used to solve the
temperature equation.  More recent opacities provided by Hartmann
and Kenyon (1996) and Bell and Lin (1994) seem to be in general
agreement with the values of Pollack et al. As Nakamoto and
Nakagawa (\cite{Nakamoto94}), we further assumed
$\tau_{P}=2.4\tau_{R}$.

% At low
% values of $\tau$ the amount of energy produced internally is much
% lower than the absorbed flux from the star and the thermal
% structure at high radius is determined by the star flux and the
% flaring of the disk and not by viscous dissipation.

Since the intensity of the turbulence viscosity is determined by the
disk temperature and determines itself part of the heating the set of
nonlinear equations must be solved simultaneously in an autoconsistent
manner once a value of $\alpha$ is given. A time-forward integration
of the density surface distribution is then possible. The problem is
more straightforward in the case of the $\beta$ prescription because
$\nu$ is then independent of temperature.

%% Note that Eq.~(\ref{TEq}) assumes that heat is {\it not}
%% transported radially, a consequence of our assumptions of
%% geometrical thinness and radial optical thickness everywhere. This
%% assumption should be relatively secure in the optically thick
%% parts of the disk where $H\ll r$. In the outer parts of the disk
%% where this assumption may not hold as firmly, temperature does not
%% play a significant role in the disk evolution.

\subsubsection{Irradiation from the central star}

Stellar photons reprocessed by the accretion disk constitute a
significant contribution to the disk's thermal budget, especially
in regions where dissipation by viscous processes is small, i.e.
mostly in the external regions of the disk. Again, a proper
treatment of stellar irradiation is beyond the scope of this work,
but we attempt to capture the essential underlying physics under
our 1D approach.

At distances much larger than the stellar radius, it can be
shown that stellar irradiation implies an effective temperature that
is a function of the star's effective temperature $T_*$ and radius
$R_*$ (Adams et al., 1988; Ruden and Pollack, 1991):
\begin{equation}
T_{l_{1}}=T_{*} \left[ \frac{2}{3\pi} \Bigl( \frac{R_{*}}{r}
\Bigr)^3 + \frac{1}{2} \Bigl(\frac{R_{*}}{r} \Bigr)^2 \Bigl(
\frac{H}{r} \Bigr) \Bigl( \frac{d \ln H}{d \ln r}-1 \Bigr)
\right]^{1/4}
\label{eq:Tl}
\end{equation}
The first term inside the brackets is the contribution due to the
finite-sized star irradiating a flat disk. The second term,
proportional to $(d\ln H/d\ln r-1)$ accounts for the flaring of
the disk. We find numerically that this factor becomes significant
only for radii $r> 50$ AU. However, this factor is a significant
source of numerical instabilities. This is an interesting problem that
certainly requires further work. At present, we choose to avoid it by
imposing
\[
d\ln H/d\ln r= 9/7,
\]
which corresponds to the (approximate) equilibrium solution for a disk
whose temperature is dominated by the flaring term (Chiang \&
Goldreich 1997). We verified that this hypothesis is autoconsistent,
i.e. our disks are close to $H\propto r^{9/7}$ in regions where
the flaring term dominates.

We additionally considered that the disk is heated by its
environment which radiates at the same temperature as the
molecular cloud, $T_{l_2}=T_{cd}$. The true irradiation
temperature $T_l$ is then computed as

\begin{equation}
T_l^4=T_{l_1}^4+T_{l_2}^4
\end{equation}

\subsection{Comparison to an $\alpha$-disk model with detailed radiative transfer.}

\begin{figure}
\centering
\includegraphics[width=8.5cm]{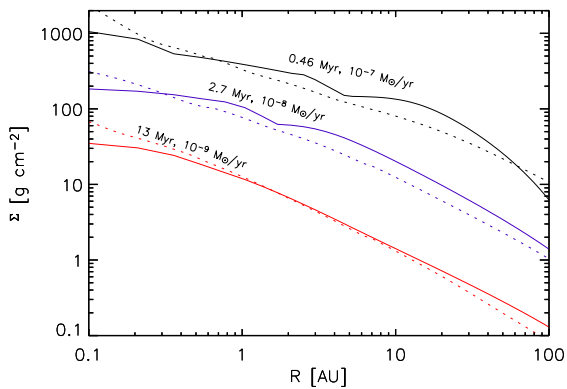}
\caption{Surface density as a function of orbital distance for
  three different times, corresponding to accretion rates onto the
  central protostar of $10^{-7}$, $10^{-8}$ and $10^{-9}\,\rm
  M_\odot\,yr^{-1}$. Plain lines show the results of our model. These
  are compared to similar calculations by D'Alessio et al. (2001) for
  a standard $\alpha$-disk with $\alpha=0.01$, $M_*=0.5\,\rm M_\odot$,
  $R_*=0.0093$\,AU and $T_*=4000$\,K (dotted lines). We further used:
  $M_{0}=0.2\,\rm M_\odot$, $\omega_{\cld}=3 \times 10^{-3}\rm\,s^{-1}$
  and $T_{\cld}=10$\,K implying a total accretion time of the cloud core
  of 0.3 Myr and an outer maximum centrifugal radius of $60$\,AU.}
\label{Fig.Dalessio1}
\end{figure}

Our model was tested against a calculation by d'Alessio et al (1999,
2001) for a 0.5\,M$_\odot$ T-Tauri star. Their calculation of the
density profile is based on a standard $\alpha$-disk model, in which a
static solution to Eq.~(\ref{Sigma}) with no source term is found by
imposing a constant mass flux throughout the disk. As shown in
fig.~\ref{Fig.Dalessio1}, the resulting surface densities for a given
accretion rate onto the star differ slightly. This is not unexpected,
due to the fact that we included the collapse phase, because assumed
opacities are different, and because the static solution is not
necessarily a good estimate of the disk's structure, especially at
large orbital distances.

\begin{figure}
\centering
\includegraphics[width=8.5cm]{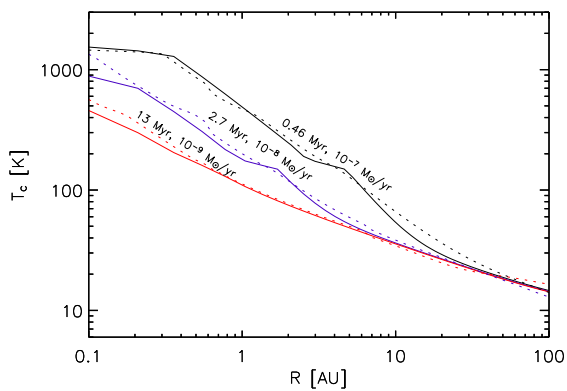}
\caption{Midplane temperatures as a function of orbital distance
  calculated in this work (plain lines) and by D'Alessio et al. (2001)
  (dotted lines), in the same conditions as in
  Fig.~\ref{Fig.Dalessio1}.}
  \label{Fig.Dalessio2}
\end{figure}

Despite these differences, the midplane temperatures inferred in our
model compare well to the more elaborate 2D calculations by D'Alessio
et al. Figure~\ref{Fig.Dalessio2} shows that the comparison is good in
two regimes: (i) In the inner parts of the disk ($R \wig{<} 10$\,AU),
where the release of gravitational energy by viscous dissipation
dominates over the star's irradiation and governs the disk's
temperature structure; (ii) In the intermediate and outer parts of the
disk, where irradiation from the central star is the dominant source
of heating and the temperature profile tends towards $T(r)\propto
r^{-1/2}$.

\subsection{Limitations of the model}

\subsubsection{Structure of the inner disk}

In the calculations that are presented, we assume that the disk
extends from an arbitrary inner boundary at 0.1 AU and beyond. The
exact location of this inner boundary is of little importance for
the evolution of the outer disk (say, beyond 10\,AU). However, the
exact structure of the inner disk and the fact that recent
observations seem to point to a clearing of the region inside $\sim
5$\,AU (Bergin et al. 2004) can affect the constraints derived from
the accretion rate. For example, it can be noticed that the presence
of a Jupiter-mass planet would lower the accretion rate onto the
star, because part of the material from the outer disk would be
accreted by the planet (e.g. D'Angelo et al. 2002, Bate et al.
2003), and part of the disk's gravitational energy would be used to
push the planet inward. Clearly, in the absence of detailed models
of this and because of other possible explanations
(photoevaporation, grain growth) it would be unrealistic to try to
also fit the observed structure of the inner disk with our simple
model. Last but not least, the error bar on $\dot M$ is relatively
large and is presumably larger than this source of uncertainty.

\subsubsection{Thermal vertical structure}

In models with an $\alpha$-viscosity, the temperature determines
the sound speed $c_s$, the scale height $H$ and hence the
viscosity $\nu$ itself. We use the midplane temperature because
the gravity near the midplane is low ($\sim \omega z$) and
vertical temperature variations should be small in these regions.
However, radiative transfer models (Malbet \& Bertout 1991; Chiang
\& Goldreich 1997; D'Alessio et al, 1999, 2001) show that in
regions where the stellar irradiation is significant, the
temperature in the outer layers of the disk is expected to be
larger than at the midplane. This was verified by (Dartois et al.,
2003) who found that in DM Tau's outer disk, the central
temperature is $T_{c}\sim 15\,$K while $T_{\rm photosphere}\sim
30\,$K.

Another aspect of the problem is that in the tenuous atmosphere of
the outer disk, the gas may not be in thermal equilibrium and may
therefore be heated more than the dust disk, i.e. by absorption of
FUV photons. Altogether, this should introduce a maximum of a factor
of 2 indetermination on the temperatures and $\alpha$-viscosities of
the outer optically thin disk.

\subsubsection{Non-viscous processes affecting the disk evolution}

Our model assumes that the present disks around DM Tau and GM Aur are
solely the result of the collapse of a molecular cloud core and of the
subsequent viscous evolution of the disk. May these disks be altered
by other physical processes? Following Hollenbach et al. (2000), we
identify five possible sources of disk dispersal:
\begin{itemize}
\item Wind stripping;
\item Photoevaporation due to an external source;
\item Photoevaporation due to the central star;
\item Tidal stripping due to close stellar encounters;
\item Planet formation.
\end{itemize}

Wind stripping may indeed affect the disk thickness, but according
to Hollenbach et al. (2000), the associated timescales for disk
dispersal are, in the regions of interest, well over 10\,Myr and
can be neglected.

Unlike in Orion (e.g. O'Dell et al., 1993), photoevaporation due
to external sources should not be a concern in the Taurus-Auriga
association due to the small number of massive stars present.

Photoevaporation due to the central star may be estimated from the
work of Hollenbach et al. (1994; see also Shu et al. 1993). The
authors estimate that the total mass loss due to UV
photoevaporation of disk material is:
\begin{equation}
\dot{M}_{UV}=4.1 \times 10^{-10}\Phi_{41}^{1/2} M_{*}^{1/2} {\rm M}_\odot
\mathrm{yr}^{-1},
\end{equation}
where $\Phi_{41}$ is the ionizing photon luminosity of the central
star in units of $10^{41}$ s$^{-1}$.  There is no precise
  value of this parameter, but Alexander et al. (2005) advocate using
  the observed He\,II/C\,IV line ratio to derive the hardness of the
  ultraviolet spectrum and obtain $\Phi_{41}\approx 1-1000$ for five
  classical T-Tauri stars. As shown by Clarke et al. (2001) and
Matsuyama et al. (2003), photoevaporation alters significantly the
inner disk's structure when $\dot{M}_{UV}\sim \dot{M}_*$, where
$\dot{M}_*$ is the accretion rate onto the star. This implies that
  photoevaporation may play a role even in DM Tau and GM Aur if the UV
  photon rate is in the upper range of that estimated by Alexander et
  al. Even in that case however, we expect the effect on the structure
  of the outer disks to be relatively modest.

Tidal stripping may be important for the evolution of accretion
disks in relatively dense environments.  A simple estimate for the
timescale of truncation of a disk to a radius $r_d$ assumes that
the disk is stripped to about 1/3 the impact parameter (Clarke and
Pringle, 1991, Hollenbach et al. 2000); hence: $t_{\rm SE}\approx
1/(n_\star \sigma v)$, where $n_\star$ is the density of stars,
$\sigma\approx \pi(3r_d)^2$ is the collision cross section and $v$
is the velocity dispersion of stars. In our case, assuming $t_{\rm
SE}\wig{<}10^7\,$yr, $v\approx 1\rm\,km\,s^{-1}$ and $r_d\approx
1000\,$AU implies that stellar encounters could become important
for star densities $n_\star\wig{>} 50\rm\,pc^{-3}$. Values
$n_\star\approx 100\,\rm pc^{-3}$ appear to be typical of the
central Taurus cloud (Clarke et al. 2000). Moreover, both DM Tau
and GM Aur are relatively isolated in this cloud and consequently
this mechanism has probably not affected their disks. This is
probably unlike other young stars, which may explain why young
stars with very extended disks remain relatively rare.

Last but not least, planet formation is a mechanism susceptible of
greatly modifying the structure of disks, particularly when {\it
giant} planets form. However, planets are thought to form closer
to the central star than can be probed by the present
observational techniques. Their effect on the disks can be thought
as a surface density sink, and should hence be relatively modest at
large orbital distances.

\subsubsection{Gravitational stability of the disk}

The local gravitational stability of a rotating disk against
axisymmetric perturbations is measured by the Toomre-$Q$
parameter, which is defined by
\begin{equation}
\label{Q-Toomre}
Q=\frac{k c_{s}}{\pi G \Sigma},
\end{equation}
where $k$ is the epicyclic frequency given by
$k^{2}=(1/r^{3})[\partial(r^{4} {\Omega_{k}}^{2})/\partial r]$
(Toomre, 1964; Goldreich and Lynden-Bell, 1965). Disks are
gravitationally stable if $Q>1$ all over the disk (the rotation
gradient dominates and the disk follows axisymmetric evolution) and
are unstable if $Q \leq 1$ anywhere (gravitation may dominate locally
and break out the axisymmetry). In this case disks may produce a
direct gravitational collapse of particles into planetesimals
(Goldreich and Ward, 1973), giant planet cores (Boss, 2003) or may
develop spiral density waves able to redistribute the angular momentum
in a non local way until stability is reached again (Laughlin and
R\'o$\dot{\textrm{z}}$yczka, 1996; Laughlin et al. 1997, 1998).  This
last case is the generally accepted scenario for the evolution of
massive disks.

Nakamoto and Nakagawa (\cite{Nakamoto94}) showed that disk instabilities
are more likely to appear in disks with low values of $\alpha$ and
high values of $\omega_{\cld}$. Laughlin and R\'o$\dot{\mathrm{z}}$yczka
(1996) found that in most of the situations the global transport of
angular momentum by spiral arms in gravitationally unstable disks
may be roughly in agreement with axisymmetric models with values
of $\alpha_s$ of $0.02-0.03$.

Although both DM Tau and GM Aur disks are at present
gravitationally stable ($Q > 1$ everywhere), they may have gone
through an early unstable phase. In order to account for this
effect, we adopted $\alpha={\rm Max}(0.03,\alpha)$ when the
condition $Q <1$ was met somewhere in the disk. A limited number
of models with $\alpha\ge 0.03$ went through an unstable phase
without any artificial increase in their viscosity. We did not
modify the viscosity of $\beta$-models, whatever the value of $Q$.

We observed {\it a posteriori} that the gravitational instability
phase was generally limited in time to less than 0.1\,Myr. An
incorrect handling of gravitational instabilities hence yields a
negligible error on the age of the system. An effect that is
potentially more significant, and ignored in the present study,
concerns the fact that the advection of disk material can depart
from a diffusive solution. To the extreme, this process may lead
to the formation of stellar or planetary companions. Clearly, a
more consistent treatment of gravitational instabilities would be
desirable.

\section{Fitting the observations}
\subsection{Setting of the calculations}

The equations described in the previous section are solved
numerically using an explicit finite-difference scheme. Each
calculation begins at time $t=M_{0}/\dot{M}$. The inner disk
boundary is set to $R_{in}=0.1$ AU, and the maximum computational
space ends at $R_{out}=10^{4}$ AU. In order to solve numerically
Eq.  (\ref{Sigma}) we follow the method described by Bath and
Pringle (1981). This requires the radial points to be equally
spaced on a $X=\sqrt{r}$ space. The stability condition required
to solve Eq. (\ref{Sigma}) is that the time-step $\Delta t$ obeys

\begin{equation}
\Delta t \geq \rm{Min} \left( \frac {X^2\Delta X^2}{24\nu} \right)
\end{equation}

An adaptive time-step scheme taking into account this condition is
used. A grid resolution of 250 points is adequate for disks with a
centrifugal radius that is always larger than 4\,AU. To calculate the
evolution of disks with smaller values of $j_{\cld}$ so that $2\wig<
\rcrit\wig< 4$\,AU, we use 500 points, corresponding to a resolution in
the inner regions $\sim 0.15\,$UA.

The model parameters are:
\begin{itemize}
\item $\alpha$ or $\beta$, which characterize the magnitude and
  functional form of the turbulent viscosity;
\item $\omega_{\cld}$, the angular velocity of the molecular cloud
  core;
\item $T_{\cld}$, the temperature of the molecular cloud (which is
  equivalent to a characteristic accretion velocity from the molecular
  cloud onto the circumstellar disk);
\item $M_0$, the initial mass of the protostar;
\item $M_{\cld}$, the total amount of material initially in the
  molecular cloud and eventually in the star+disk system.
\end{itemize}

These parameters reflect either uncertainties in the physics of star
formation ($\alpha$-$\beta$, $M_0$) or unknown initial conditions
($\omega_{\cld}$, $M_{\cld}$), or both ($T_{\cld}$ through the accretion
rate of the molecular cloud, see Eq.~\ref{eq:mdot}).

Our approach is to constrain these parameters by extensive
numerical simulations and comparison with observations. Four sets
of calculations are performed for DM Tau and GM Aur, using
either the $\alpha$ or $\beta$ parameterizations and considering
the Shu (1977) classical cloud collapse model. To explore the
sensitivity of the results to the cloud collapse model, we calculate a
fifth set of model for the DM Tau $\alpha$ case and the
collapse scenario derived from Basu (1998).

\begin{table}
         \caption[]{Explored parameter space}
         \begin{tabular}{p{3.2cm} p{2.0cm} p{2.0cm}lcc}
            \hline
            \vspace{0.005in}                  \smallskip     &
            \vspace{0.005in} \textbf{DM Tau}  \smallskip     &
            \vspace{0.005in} \textbf{GM Aur}  \smallskip      \\
            \hline
            $M_{0}\rm\ [M_\odot]$          & 0.05 and 0.30         & 0.05 and 0.40          \\
            $M_{\cld}\rm\ [M_\odot]$         & $0.4-1.0$             & $0.4-1.5$              \\
            $\omega_{\cld}\rm\ [s^{-1}]$     & $10^{-15}-10^{-11}$   & $10^{-15}-10^{-11}$    \\
            $T_{\cld}\ [K]$                  & $3.0-30.0$            & $3.0-30.0$             \\
            Accretion Time($^{*}$) [yr]    & $45000-5\times 10^6$  & $49000-5\times 10^6$   \\
            $\alpha$            & $10^{-5}-1$   & $10^{-5}-1$        \\
            $\beta$             & $10^{-6}-0.1$ & $10^{-6}-0.1$      \\
            Number of models    & 2000 $\alpha$ & 2000 $\alpha$      \\
                                & 1500 $\beta$  & 1500 $\beta$       \\
            Magnetic cloud core & 2000 $\alpha$ &                    \\
            \hline
        \label{Parameters}
        \end{tabular}
        ($^{*}$) Here the accretion time refers to the time it takes
        for the molecular cloud core to entirely collapse onto the
        disk (see Eq.~\ref{eq:mdot}).
\end{table}

In order to efficiently explore the space of parameters we use a
Monte-Carlo procedure to chose the values of the model parameters
within the range given in table~\ref{Parameters}.  Two values of
$M_{0}$ are chosen. A small value is representative of cases in which
the central condensation grows relatively slowly, and gravitationnal
instabilities dominate the early disk evolution.  A large value is
representative of an initially evolved situation in which the
proto-star forms rapidly, and the disk grows significantly only at
later times. This could occur if ambipolar diffusion and jets allow an
efficient outward transport of angular momentum early on during the
collapse of the molecular cloud.

Cases for which the initial centrifugal radius $\rcrit$ (see
Eq.~(\ref{Rd})) is larger than $\sim$5000\ AU are disregarded, as
well as those with final centrifugal radius too small to solve
accurately the building-up of the disk.  For numerical reasons
models that require a time-integration with a step smaller than
0.5 yrs are not calculated. Table 2 provides a summary of the
different sets of calculations run and the coverage of the space
of parameters. Table~\ref{ParametersLimits} provides the numerical
limitations imposed by the two different grid resolutions, the
minimum time-step required, and the resulting limits on $j_{\cld}$
and $\nu_0$. The values of $j_{\cld}$ bracket inferred values for
Class~I protostellar envelopes,
$10^{20}\wig{<}j_{\cld}\wig{<}10^{21}\rm\,cm^2\,s^{-1}$ (Ohashi et
al. \cite{Ohashi97}). The upper limit on $\nu_0$ is relatively
large and limits us only for extreme values of the
$\beta$-parameter in the case of DM Tau (see
figure~\ref{FgViscosities} hereafter).

\begin{table}
        \caption[]{Numerical limitations of the calculation}
        \begin{tabular}{p{3.9cm} p{3.9cm} lc}
        \hline
        \vspace{0.005in} \textbf{$N_{points}=$250}  \smallskip   &
        \vspace{0.005in} \textbf{$N_{points}=$500}  \smallskip     \\
        \hline
        $\Delta R_{0}=0.4$                & $\Delta R_{0} = 0.15$             \\
         $2.3   <  \rcrit           < 5000$  &  $0.8  < \rcrit             < 5000$  \\
         $19.0  <$ Log$_{10}$ $(j_{\cld}) < 21$    &  $18.8 < $ Log$_{10}$ $(j_{\cld}) < 21$     \\
        \hline
        \vspace{0.005in}
         $\Delta t$ $> 0.5$  $\Rightarrow$ \\
         Log$_{10}$ $(\nu_0) < 17.2$    &  Log$_{10}$ $(\nu_0) < 16.5$       \\
        \hline
        \label{ParametersLimits}
        \end{tabular}
        $\Delta R_{0}$ is the spatial resolution of the model at the inner disk boundary
        and $\rcrit$ is the centrifugal radius, both are given in AU,
        $j_{\cld}$ is the specific angular momentum of the molecular cloud
        and is given in cm$^{2}$\,s$^{-1}$, $\Delta t$ is the minimum allowed
        computational time-step in yr and is related with $\nu_0$ the value of
        the disk viscosity at the inner boundary given in cm$^{-2}$\,s$^{-1}$.
        The inner disk boundary is located at $0.1$\,AU.
\end{table}

The calculations were performed on an 8PCs Linux/Beowulf cluster
over several months. The low $\Delta t$ required by the stability
criterion imply total computation times for each model that range
between minutes to several hours for the most computationally
expensive models (specifically, these correspond to 500-points,
$\beta$ models of GM Aur). The large database of resulting models
is then compared to the various observational constraints in order
to extract the possible values of the main physical parameters.

\subsection{Selecting models from observational constraints}
\label{section:select}

\begin{figure}
\includegraphics[width=8.5cm]{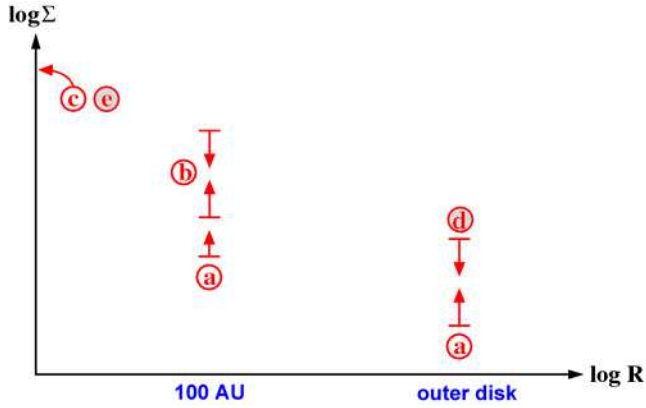}
\caption{Illustration showing the different observational constraints
  used in this work, in terms of surface density and orbital
  distance. The labels ``c'' and ``e'' correspond to constraints on
  the accretion rate onto the central star. Constraints ``e'' and ``d'' are reasonable expectations
  of the mass accretion rate and outer disk maximum surface density.}
\label{SchemeSelection}
\end{figure}

The observational constraints characterizing DM Tau and GM Aur
have very different origins. Three independent constraints are used
(see \S~2):
\begin{description}
\item[(a)] The fact that the disks are optically thick in
  $^{12}$CO at 100\, AU and the outer disk radius requires that
  $\Sigma$ be larger than a minimal value. This value is calculated
  by assuming that all C is in form of vapor CO and translates
  into a minimal value of $\Sigma$ at $R_{out}$.
\item[(b)] Values of $\Sigma$ at 100 AU have to be consistent with
  those inferred from dust emission, allowing for uncertainties in the
  opacity coefficient, grain size and dust to gas ratio.
\item[(c)] The accretion rate onto the star has to fit constraints
  obtained from the star's excess luminosity. A conservative error bar
  of 0.54 in Log$_{10}(\dot{M})$ is assumed.
\end{description}
In order to further narrow the possible solution ensemble, we
choose to add the following reasonable constraints:
\begin{description}
\item[(d)] At the outer $^{12}$CO radius, the value of $\Sigma$
cannot be larger than 20 times the lower limit. This value is
inspired by the discrepancy between the CO and dust emission
observations at 100\,AU.
\item[(e)] The inferred accretion rate is
supposed to be known within a realistic but model dependent error
bar of only 0.3 in Log$_{10}(\dot{M})$.
\end{description}
These constraints are schematically shown on
Figure~\ref{SchemeSelection}. They should be further refined by
combining observational and theoretical studies. In this work, we
purposely use rather unrestrictive interpretations of the observations
to obtain robust constraints on the desired physical parameters.

Together with the constraints on the star's mass and age, we thus
derive five sets of solutions, from weaker to stronger
constraints:
\begin{description}
\item[\{1\}]: (a)
\item[\{2\}]: (a)+(b)
\item[\{3\}]: (a)+(b)+(c)
\item[\{4\}]: (a)+(b)+(c)+(d)
\item[\{5\}]: (a)+(b)+(c)+(d)+(e)
\end{description}
We will mostly discuss the results obtained with sets \{4\} and \{5\}.

\subsection{Examples of models behavior: Disk history and surface density-temperature evolution}
\label{sec:Generalbehavior}

Before analyzing our global results, it is useful to describe the
general behavior of the models in terms of two specific examples. We
explain here how the observational constraints
are used to select a model as a successful fit to the data. We
consider two examples based on an $\alpha$-model of DM Tau. The parameters
for both examples are listed in Table~\ref{TbExamples}

\begin{table}
         \caption[]{Model parameters for examples 1 and 2}
        \begin{tabular}{p{2.5cm} p{1.7cm} p{1.7cm} p{1.1cm} l|c|c|c}
        \hline
        \vspace{0.005in} \textbf{Parameters}   &
        \vspace{0.005in} \textbf{Example 1}     &
        \vspace{0.005in} \textbf{Example 2}     &
        \vspace{0.005in} \textbf{Units}          \\
    \vspace{0.005in} (fixed) & & & \\
        \hline
        $\alpha$              & 0.01                 &   0.025                & --- \\
        $\omega_{\cld}$         & $2.3\times 10^{-14}$ &   $2.6\times 10^{-13}$ & (s$^{-1}$) \\
        $T_{\cld}$              & 14                   &   17 & (K) \\
        $M_{0}$               & 0.05                 &   0.05 & (M$_{\odot}$)\\
        $M_{\cld}$              & 0.515                &   0.585 & (M$_{\odot}$) \\
        \hline
        \end{tabular}
        \begin{tabular}{p{8.7cm} l}
        \vspace{0.005in}  (derived)   \\
        \end{tabular}
        \begin{tabular}{p{2.5cm} p{1.7cm} p{1.7cm} p{1.1cm} l|c|c|c}
        \hline
        \vspace{0.005in}  Log$_{10}$($J_{\cld}$)  &
        \vspace{0.005in}  52.3                  &
        \vspace{0.005in}  53.4                  &
        \vspace{0.005in} (g\,cm$^2$\,s$^{-1}$)        \\
        Log$_{10}$($j_{\cld}$)  & 19.3      &   20.3 & (cm$^2$\,s$^{-1}$)\\
        $\rcrit$               & 11        &   830  & (AU) \\
        Accretion time        & 0.18      &   0.15 & (Myr) \\
        \hline
        \label{TbExamples}
        \end{tabular}
\end{table}

In all cases, our story starts with a protostar of mass $M_{0}$, no
disk and a reservoir of cloud material with mass $M_{\cld}-M_{0}$.
Material falls from the molecular cloud at a constant rate inside a
disk of growing size $\rcrit$, a consequence of angular momentum
conservation and the inside-out collapse of a molecular cloud core in
solid rotation. Because viscous diffusion is more efficient
than the increase of $\rcrit$ with time, the disk spreads beyond the
centrifugal radius. The disk can be quite hot in this early phase,
especially when $\rcrit$ is small and the accretion rate is large
(large $T_{\rm \cld}$). The disk grows in mass until all of the
available mass in the molecular cloud core has been accreted to the
disk. After that, the central star continues to accrete disk material
while the disk expands and cools.

\begin{figure}
\includegraphics[width=8.5cm]{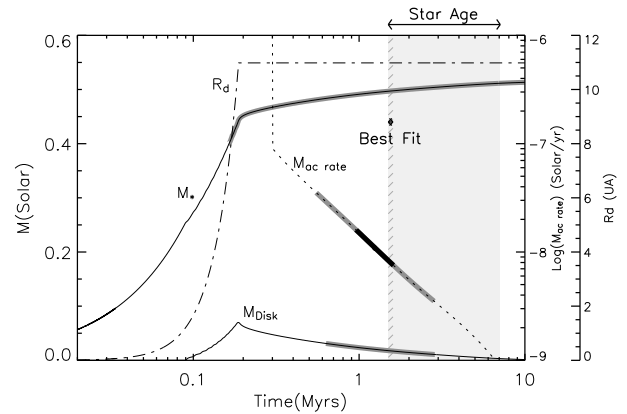}
\caption{Example 1. Evolution of star mass $M_{*}$ and disk mass
$M_{\rm disk}$ as a function of time with masses in solar units
(corresponding axis to the left) for the model parameters of
Table~\ref{TbExamples}. The accretion rate onto the
central star is shown as a dotted line (corresponding axis: first
to the right). The evolution of the centrifugal radius $\rcrit$
(see Eq.~(\ref{Rd})) is shown as a dash-dotted line (corresponding
axis: far-right). Gray curves and the hashed region indicate time
sequences when selected observational constraints are verified
(see text).}
\label{ExampleMass}
\end{figure}

\begin{figure}
\includegraphics[width=8.0cm]{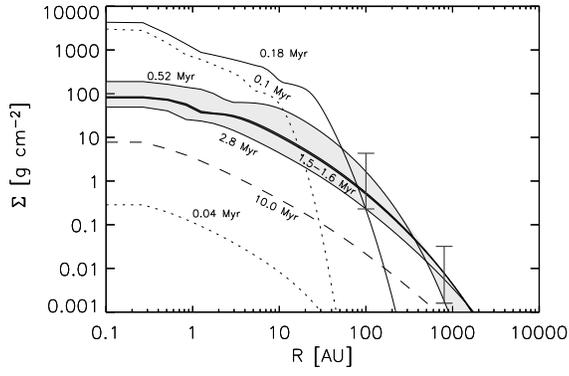}
\caption{Example 1. Surface density versus orbital distance at
different times for the model parameters of Table~\ref{TbExamples}.
Dotted lines correspond to the early formation of the disk. The
collapse of the molecular cloud ends after 0.18 Myr. The dashed line
at 10 Myr corresponds to the end of the simulation. The error bars
represent the $\Sigma$ values at 100 AU and in the outer radius that
are used as observational constraints for DM Tau. The gray area
shows the ensemble of models fitting those constraints. The
dark-shaded region shows the $\Sigma$ distribution at the time-lapse
when all observational constraints are satisfied (see text).}
\label{SigmaEvtn}
\end{figure}

\begin{figure}
\includegraphics[width=8.5cm]{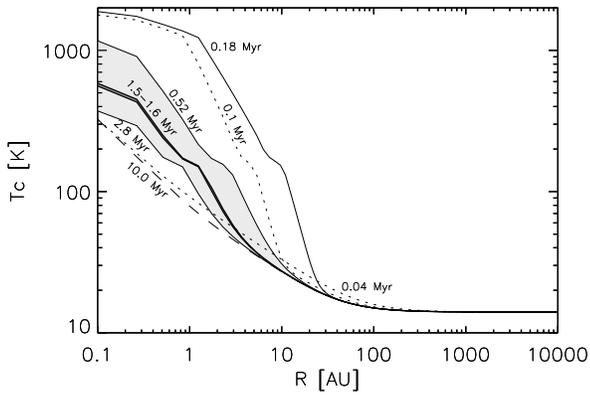}
\caption{Example 1. Same as fig.~\ref{SigmaEvtn}, but for the
  midplane temperature as a function of orbital distance.}
\label{DiskTemp}
\end{figure}

Starting with Example~1, fig.~\ref{ExampleMass} illustrates the
evolution of the star mass, disk mass, centrifugal radius and star
accretion rate.  Figure~\ref{SigmaEvtn} shows the evolution of the
surface density of the disk. The disk construction phase (dotted
lines) appears as a quick phase in which the disk density
increases by orders of magnitude with a relatively sharp outer
edge. When the molecular cloud has collapsed entirely (at
$t=0.18$\,Myr), the disk has already spread further than the
maximum centrifugal radius (11 AU), a consequence of the
relatively large $\nu$. The later evolution of the system is
characterized by the inner accretion of disk material and the
outer diffusion of angular momentum. After 3 Myr, the diffusion of
material proceeds much more slowly due to the lower density,
temperature and viscosity of the inner disk.

The midplane temperatures are shown in Fig.~\ref{DiskTemp} for the
same time-sequence. The disk heats up considerably during the
molecular cloud collapse phase (the first 0.2 Myrs) and cools down
gradually when the collapse ceases. The sharp transitions in the
temperature curves arise from the different regimes of Rosseland
opacities at different temperatures (Ruden and Pollack, 1991). The
outer disk beyond 50 AU is optically thin, vertically extended,
flared, and its thermal structure is determined solely by stellar
irradiation and a diffuse heat source of $T_{cd}$.

In Example~1, the disk is always stable to gravitational
perturbations ($Q>1$) and the disk mass is never a large fraction
of the star mass. This model evolves smoothly with values of the
initial parameters in good agreement with expected values in the
Taurus Aurigae region. The question then is: Does this model
satisfy the observational requirements for DM Tau?  And in that
case, what other values of the set of parameters representative
perhaps of very different initial conditions or turbulence in the
disk, would also agree with the observations?

Figure~\ref{ExampleMass} shows thick grey lines superimposed on each
plotted quantity. Each one shows the range of time for which that
quantity agrees with the available observations. For $M_{Disk}$, the
grey line represents the period of time when the $\Sigma$ surface
density satisfies the observational error bars discussed in
Section~2. This period is easily identified on
Figure~\ref{SigmaEvtn}. The accretion rate is reproduced either
within large error bars (thick grey line), or with small error bars
(thick black line) (see Section~\ref{sect:Obs}). The uncertainty
over the star age for DM Tau is marked as a light-grey box.
Figure~\ref{ExampleMass} hence shows that the model is a good fit to
the data from 1.5 to 2.8 Myrs (surface densities and star age), from
1.5 to 2.6 Myrs ($\dot{M}$ with its large error) or from 1.5 to 1.6
Myrs (small error bars on $\dot{M}$). The latter is shown as a
dashed region.

This model hence does fulfill the ``strict'' observational
constraints (set \{3\}) and also the reasonable additions (sets
\{4\} and \{5\}). This example shows how DM Tau's 800\,AU disk can
be formed by viscous diffusion of an initially much smaller disk,
with a centrifugal radius $\rcrit$=11 AU.

\begin{figure}
\includegraphics[width=8.5cm]{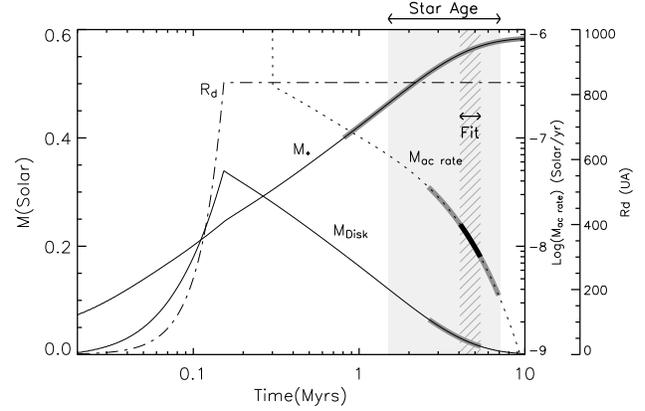}
\caption{Example 2. Evolution of star mass $M_{*}$ and disk mass
$M_{\rm disk}$ as a function of time. See fig.~\ref{ExampleMass} and
  Table~\ref{TbExamples} for details.} \label{ExampleMass2}
\end{figure}

\begin{figure}
\includegraphics[width=8.0cm]{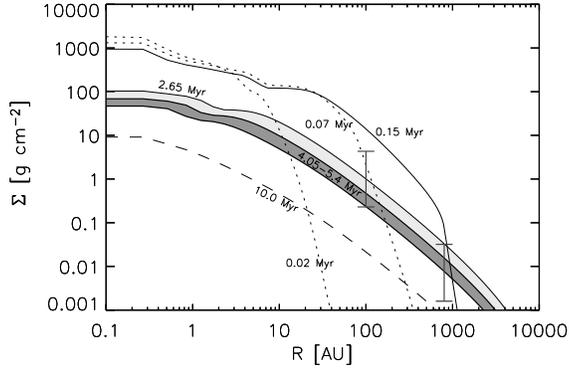}
\caption{Example 2. Surface density versus orbital distance at
different times. See fig.~\ref{SigmaEvtn} and
  Table~\ref{TbExamples} for details.} \label{SigmaEvtn2}
\end{figure}

\begin{figure}
\includegraphics[width=8.5cm]{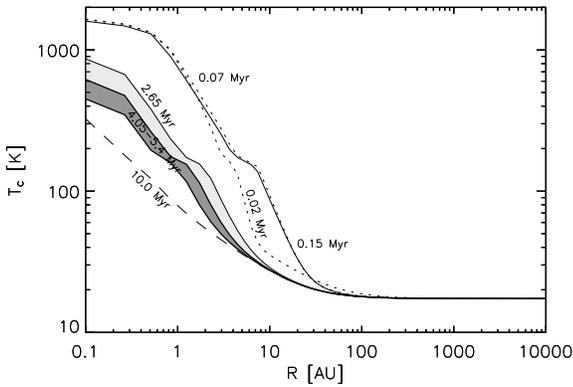}
\caption{Example 2. Midplane temperature versus orbital distance. See
  fig.~\ref{DiskTemp} and Table~\ref{TbExamples} for details.}
  \label{DiskTemp2}
\end{figure}

Let us now examine our second example. Here the disk is assumed to
form in 0.15 Myrs with a maximum centrifugal radius of 800 AU.
Most of the disk material falls so far from the central star that
the disk gets more massive than the central star at the end of the
collapse (figs.~\ref{ExampleMass2} and \ref{SigmaEvtn2}). Yet, the
disk diffuses outwards and gets accreted into the central star.
For a certain period of time (4-5.4 Myrs) it also satisfies all
the observational constraints we have considered.

Observationally it may be very difficult to distinguish between
these two scenarios at a late and evolved age but each one has
very different implications for the formation of planetary
systems. In the first case, a massive, relatively small disk
forms. This may allow a rapid build-up of planetesimals and
protoplanetary cores, but may induce a fast inward migration. In
the second example, planetesimals could still be formed in an
extended disk on longer timescales. If massive cores were to be
formed late in the system their inner migration could be damped.
The possibility of forming planetesimals and planets in this
variety of scenarios will be studied in a forthcoming paper.

As we will see in the next section, solutions matching the
observed surface densities and other constraints are far from
being limited to these two cases; a variety of initial conditions
and values of the turbulent viscosity can yield satisfactory
models. We now turn to an analysis of the ensemble of models that
match the observations.

\section{Results: Allowed models and derived theoretical constraints}
\label{section:Results}

\subsection{General}
\label{section:GeneralResults}

For the most of this section we discuss results on the basis of only 2
parameters representative of the disk physics ($\alpha$ or $\beta$)
and of the initial conditions ($j_{\cld}$), respectively. We thus
derive constraints on these parameters depending on the chosen set of
observational constraints. Since given values of ($\alpha$ or $\beta$,
$j_{\cld}$) may be obtained from different combinations of the initial
conditions, we also plot the fraction of models that fit the
observations.  We further checked that the global results are not
dependent on the value of $M_{0}$ or on the numerical resolution.

\begin{figure*}
\centering
\includegraphics[width=17.0cm]{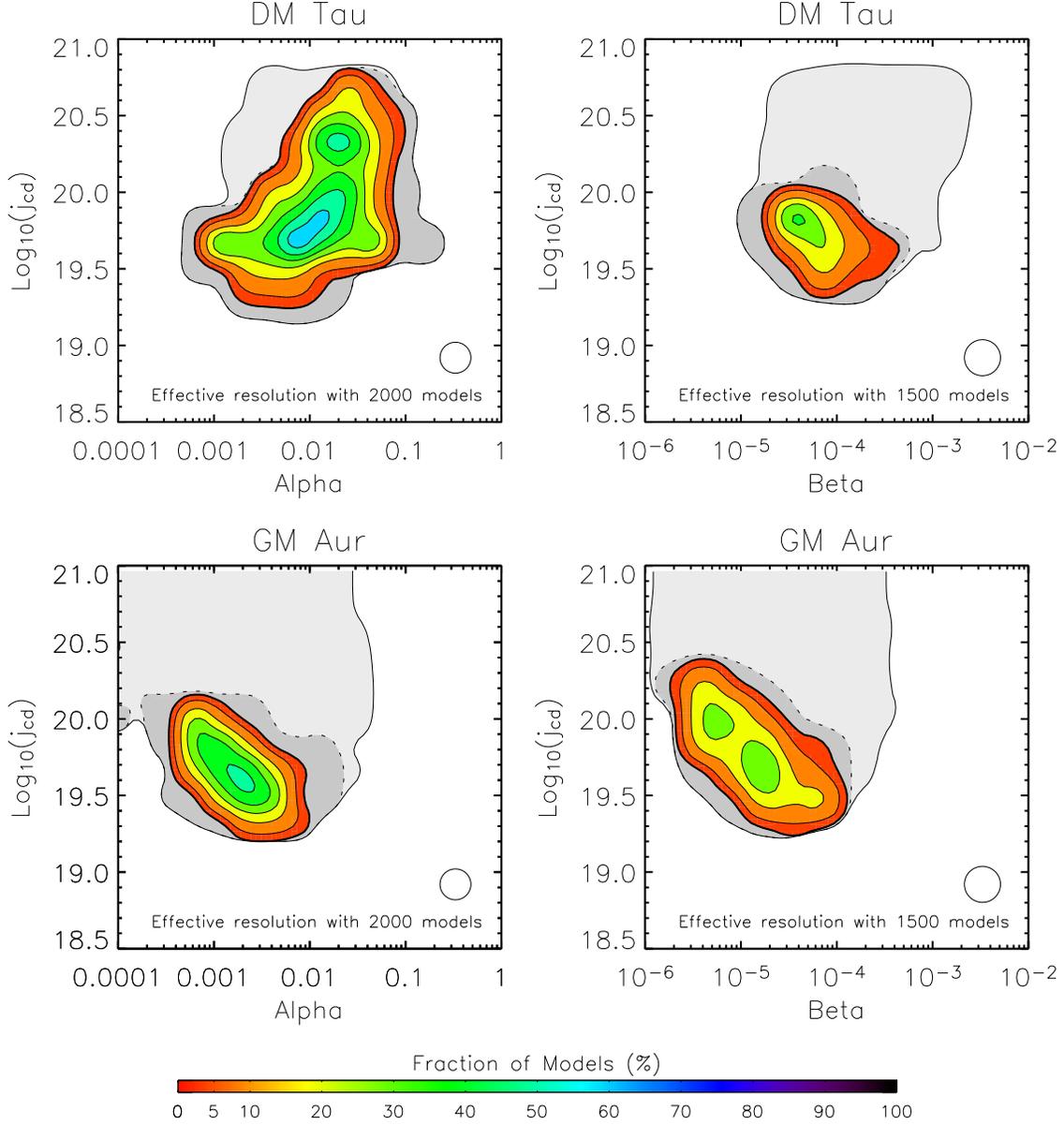}
\caption{Viscosity and initial specific angular momentum of models
fitting DM Tau and GM Aur for both parameterizations of
turbulence. The contour colored plot represents the fraction of
models fitting the strongest observational constraints (set
\{5\}). The light-shaded region around, contoured by a dotted
line, represents the additional area covered by models assuming an
uncertain accretion rate (set \{4\}). The light grey filled region
surrounding this area corresponds to models with no upper limit on
the surface density in the disk's outer region (set \{3\}).The
spatial resolution for each panel is shown as an empty circle. The
circle in each figure shows the approximate spatial resolution of
the figure obtained by the Monte-Carlo procedure for an accuracy
in the fraction of models of 10\%. Features in the plots smaller
than each circle are not well resolved.}
\label{ParameterSpace}
\end{figure*}

\begin{table*}[htb]
\begin{tabular}{p{0.5cm}p{0.40cm}p{0.5cm}p{0.4cm}p{13.7cm}}
\hline \vspace{0.005in}\textbf{System}     &
 &
\vspace{0.005in}\textbf{Param.}  &
 &
\vspace{0.005in}\textbf{Model results obtained for the different observational constraints} \smallskip \\
\hline
\end{tabular}

\begin{tabular}{p{0.7cm}p{0.1cm}p{0.6cm}p{0.4cm}
                p{2.9cm}p{2.9cm}p{3.4cm}p{3.7cm}}
 &   &   &  & \textbf{\{2\}:} CO + Dust
            & \textbf{\{3\}:} \{2\} + $\dot{M}$
            & \textbf{\{4\}:} \{3\} + Outer Limit
            & \textbf{\{5\}:} \{4\} + Accurate $\dot{M}$   \\
\end{tabular}

\begin{tabular}{p{1.20cm}p{0.0cm}p{0.45cm}p{0.0cm}
                p{0.7cm}p{0.8cm}p{0.8cm}    %  CO + Dust
                p{0.7cm}p{0.8cm}p{0.8cm}    %  CO + Dust + Ac. Rate
                p{0.7cm}p{0.8cm}p{0.8cm}    %  CO + Dust + Ac. Rate + Out limit
                p{0.7cm}p{0.8cm}p{0.8cm}}   %  CO + Dust + Ac. Rate + Out limit + Strong AC.RATE
 &   &   &           &
    Min &  Mean &  Max &
    Min &  Mean &  Max &
    Min &  Mean &  Max &
    Min &  \textbf{Mean} &  Max \\
\hline
  DM Tau  &   & $\alpha$      &               &
5x$10^{-4}$ & 0.03  & 0.6      &   % CO + Dust
5x$10^{-4}$ & 0.02  & 0.15       &   % CO + Dust + Ac. Rate
5x$10^{-4}$ & 0.02  & 0.15       &   % CO + Dust + Ac. Rate + Out limit
8x$10^{-4}$ & \textbf{0.02}  & 0.08      \\  % CO + Dust + Ac. Rate + Out limit + Strong AC.RATE
          &   & $j_{\cld}$&                &
  19.0     &20.1    & 20.8      &   % CO + Dust
  19.2     &20.1    & 20.7      &   % CO + Dust + Ac. Rate
  19.2     &20.0    & 20.7      &   % CO + Dust + Ac. Rate + Out limit
  19.3     &\textbf{20.0} & 20.7     \\  % CO + Dust + Ac. Rate + Out limit + Strong AC.RATE
           &   & $\rcrit$&               &
  2.4      &1090   & 4900       &   % CO + Dust
  4.6      &980    & 4900       &   % CO + Dust + Ac. Rate
  4.6      &720    & 4900       &   % CO + Dust + Ac. Rate + Out limit
  7.3      &\textbf{750}    & 4900      \\   % CO + Dust + Ac. Rate + Out limit + Strong AC.RATE
\hline
           &   & $\omega_{\cld}$&                &
    0.14   &  22   & 218        &   % CO + Dust
    0.14   &  23   & 218        &   % CO + Dust + Ac. Rate
    0.14   &  22   & 218        &   % CO + Dust + Ac. Rate + Out limit
    0.14   &  \textbf{23}   & 218        \\  % CO + Dust + Ac. Rate + Out limit + Strong AC.RATE
           &   & $T_{\cld}$&                &
     2     &  15 & 30         &   % CO + Dust
     2     &  16 & 30         &   % CO + Dust + Ac. Rate
     2     &  16 & 30         &   % CO + Dust + Ac. Rate + Out limit
     2     &  \textbf{16} & 30         \\  % CO + Dust + Ac. Rate + Out limit + Strong AC.RATE
           &   & $M_{\cld}$&                &
  0.41     & 0.59 & 1.0       &   % CO + Dust
  0.41     & 0.56 & 0.87      &   % CO + Dust + Ac. Rate
  0.41     & 0.53 & 0.69      &   % CO + Dust + Ac. Rate + Out limit
  0.41     & \textbf{0.53} & 0.69      \\  % CO + Dust + Ac. Rate + Out limit + Strong AC.RATE
           &   & $J_{\cld}$&                &
  52.0     &53.5   & 54.1      &   % CO + Dust
  52.2     &53.1   & 54.0      &   % CO + Dust + Ac. Rate
  52.2     &53.0   & 53.9      &   % CO + Dust + Ac. Rate + Out limit
  52.3     &\textbf{53.0}   & 53.7      \\  % CO + Dust + Ac. Rate + Out limit + Strong AC.RATE
           &   & AT&                    &
  0.05     &0.5    & 4.4        &   % CO + Dust
  0.05     &0.4    & 4.0        &   % CO + Dust + Ac. Rate
  0.05     &0.3    & 4.0        &   % CO + Dust + Ac. Rate + Out limit
  0.05     &\textbf{0.3 }   & 3.7        \\  % CO + Dust + Ac. Rate + Out limit + Strong AC.RATE
           &   & Age&                    &
  1.5      &3.7    & 7.0        &   % CO + Dust
  1.5      &3.9    & 7.0        &   % CO + Dust + Ac. Rate
  1.5      &3.5    & 7.0        &   % CO + Dust + Ac. Rate + Out limit
  1.5      &\textbf{3.5}    & 6.8        \\  % CO + Dust + Ac. Rate + Out limit + Strong AC.RATE
           &   & $\Delta$Age     &     &
  0.05     &2.6    & 5.5        &   % CO + Dust
  0.05     &2.3    & 5.5        &   % CO + Dust + Ac. Rate
  0.05     &2.0    & 5.5        &   % CO + Dust + Ac. Rate + Out limit
  0.05     &\textbf{1.0}  & 2.8        \\  % CO + Dust + Ac. Rate + Out limit + Strong AC.RATE
           &   & M$_{disk}$     &     &
  0.006    &0.05   & 0.42       &
  0.009    &0.036  & 0.38       &
  0.009    &0.025  & 0.10       &
  0.014    &\textbf{0.036} & 0.097    \\
           &   & N$_{models}$   &     &
       &371    &         &   % CO + Dust
       &313    &         &   % CO + Dust + Ac. Rate
       &225    &         &   % CO + Dust + Ac. Rate + Out limit
       &\textbf{166}    &         \\  % CO + Dust + Ac. Rate + Out limit + Strong AC.RATE
\hline
  DM Tau  &   & $\beta$x$10^5$ &               &
 1.1      &  31   &  390       &   % CO + Dust
 1.1      &  24   &  160       &   % CO + Dust + Ac. Rate
 1.1      &  8.5  &   50       &   % CO + Dust + Ac. Rate + Out limit
 2.2      &  \textbf{8.5}  &  50           \\  % CO + Dust + Ac. Rate + Out limit + Strong AC.RATE
          &   & $j_{\cld}$&                &
 19.1     & 20.0  &  20.8      &   % CO + Dust
 19.3     & 20.1  &  20.8      &   % CO + Dust + Ac. Rate
 19.3     & 19.7  &  20.1      &   % CO + Dust + Ac. Rate + Out limit
 19.4     & \textbf{19.7}  &  20.0       \\  % CO + Dust + Ac. Rate + Out limit + Strong AC.RATE
           &   & $\rcrit$&               &
  5        &  752  &  4800      &   % CO + Dust
  9        &  810  &  4800      &   % CO + Dust + Ac. Rate
  9        &  74   &  328       &   % CO + Dust + Ac. Rate + Out limit
  15       &  \textbf{71 }  &  170      \\   % CO + Dust + Ac. Rate + Out limit + Strong AC.RATE
\hline
           &   & $\omega_{\cld}$&                &
  0.05     & 14    &  140       &   % CO + Dust
  0.1      & 15    &  140       &   % CO + Dust + Ac. Rate
  0.1      & 8     &  36        &   % CO + Dust + Ac. Rate + Out limit
  0.1      & \textbf{8 }    &   36       \\  % CO + Dust + Ac. Rate + Out limit + Strong AC.RATE
           &   & $T_{\cld}$&                &
   2       & 14    &  30        &   % CO + Dust
   2       & 15    &  30        &   % CO + Dust + Ac. Rate
   2       & 15    &  30        &   % CO + Dust + Ac. Rate + Out limit
   2       & \textbf{15}    &  30        \\  % CO + Dust + Ac. Rate + Out limit + Strong AC.RATE
           &   & $M_{\cld}$&                &
   0.413   &  0.65 & 0.97      &   % CO + Dust
   0.42    &  0.65 & 0.97      &   % CO + Dust + Ac. Rate
   0.42    &  0.56 & 0.72      &   % CO + Dust + Ac. Rate + Out limit
   0.43    &  \textbf{0.55} & 0.65      \\  % CO + Dust + Ac. Rate + Out limit + Strong AC.RATE
           &   & $J_{\cld}$&                &
  52.1     & 53.1  &   54.1    &   % CO + Dust
  52.4     & 53.2  &   54.1    &   % CO + Dust + Ac. Rate
  52.4     & 52.8  &   53.1    &   % CO + Dust + Ac. Rate + Out limit
  52.5     & \textbf{52.8}  &   53.1    \\  % CO + Dust + Ac. Rate + Out limit + Strong AC.RATE
           &   & AT&                    &
 0.06      &  0.7  & 4.9        &   % CO + Dust
 0.06      &  0.6  & 4.9        &   % CO + Dust + Ac. Rate
 0.06      &  0.4  & 2.6        &   % CO + Dust + Ac. Rate + Out limit
 0.06      &  \textbf{0.5 } & 2.6        \\  % CO + Dust + Ac. Rate + Out limit + Strong AC.RATE
           &   & Age&                    &
  1.5      &  4.0  & 7.0        &   % CO + Dust
  1.5      &  4.0  & 7.0        &   % CO + Dust + Ac. Rate
  1.5      &  3.2  & 7.0        &   % CO + Dust + Ac. Rate + Out limit
  1.5      &  \textbf{2.7}  & 4.5        \\  % CO + Dust + Ac. Rate + Out limit + Strong AC.RATE
           &   & $\Delta$Age     &     &
  0.05     &  2.7  & 5.5        &   % CO + Dust
  0.05     &  2.0  & 5.5        &   % CO + Dust + Ac. Rate
  0.05     &  0.8  & 3.4        &   % CO + Dust + Ac. Rate + Out limit
  0.05     &  \textbf{0.4}  & 1.3        \\  % CO + Dust + Ac. Rate + Out limit + Strong AC.RATE
           &   & M$_{disk}$     &     &
  0.01     &0.15    & 0.57       &
  0.01     &0.20   & 0.50       &
  0.01     &0.06   & 0.13       &
  0.02     &\textbf{0.07 } & 0.13    \\
           &   & N$_{models}$   &     &
     &  277 &        &   % CO + Dust
     &  235 &        &   % CO + Dust + Ac. Rate
     &  75  &        &   % CO + Dust + Ac. Rate + Out limit
     &  \textbf{44}  &      \\  % CO + Dust + Ac. Rate + Out limit + Strong AC.RATE
\hline
  GM Aur   &       & $\alpha$  &               &
2x$10^{-5}$& 0.02  & 0.3       &   % CO + Dust
2x$10^{-5}$& 0.005  & 0.03       &   % CO + Dust + Ac. Rate
1x$10^{-4}$& 0.003  & 0.02       &   % CO + Dust + Ac. Rate + Out limit
4x$10^{-4}$& \textbf{0.002}  & 0.01      \\  % CO + Dust + Ac. Rate + Out limit + Strong AC.RATE
           &   & $j_{\cld}$&     &
 19.2      &  20.2 &   20.9    &   % CO + Dust
 19.2      &  20.2 &   20.9    &   % CO + Dust + Ac. Rate
 19.2      &  19.6 &   20.1    &   % CO + Dust + Ac. Rate + Out limit
 19.2      &  \textbf{19.6} &   20.1    \\  % CO + Dust + Ac. Rate + Out limit + Strong AC.RATE
           &   & $\rcrit$&               &
      4.1  & 1000  &  4900      &   % CO + Dust
      4.1  & 915   &  4900      &   % CO + Dust + Ac. Rate
      4.1  & 45    &  233       &   % CO + Dust + Ac. Rate + Out limit
      4.1  & \textbf{36 }   &  199       \\   % CO + Dust + Ac. Rate + Out limit + Strong AC.RATE
\hline
           &       & $\omega_{\cld}$&                &
  0.04     &  12   &  125       &   % CO + Dust
  0.04     &  11   &  125       &   % CO + Dust + Ac. Rate
  0.04     &  3.7  &  16        &   % CO + Dust + Ac. Rate + Out limit
  0.04     &  \textbf{3.6}   &  12        \\  % CO + Dust + Ac. Rate + Out limit + Strong AC.RATE
           &   & $T_{\cld}$&                &
     2     &  16   & 30         &   % CO + Dust
     2     &  16   & 30         &   % CO + Dust + Ac. Rate
     2     &  17   & 30         &   % CO + Dust + Ac. Rate + Out limit
     2     &  \textbf{17}   & 30         \\  % CO + Dust + Ac. Rate + Out limit + Strong AC.RATE
           &   & $M_{\cld}$&                &
   0.55    &  1.0  & 1.5       &   % CO + Dust
   0.55    &  1.0  & 1.5       &   % CO + Dust + Ac. Rate
   0.55    &  0.8  & 1.3       &   % CO + Dust + Ac. Rate + Out limit
   0.55    &  \textbf{0.8}  & 1.1       \\  % CO + Dust + Ac. Rate + Out limit + Strong AC.RATE
           &   & $J_{\cld}$&                &
 52.3      &  53.5 &  54.4     &   % CO + Dust
 52.3      &  53.5 &  54.4     &   % CO + Dust + Ac. Rate
 52.3      &  52.9 &  53.4     &   % CO + Dust + Ac. Rate + Out limit
 52.3      &  \textbf{52.8}  & 53.3      \\ % CO + Dust + Ac. Rate + Out limit + Strong AC.RATE
           &   & AT&                    &
 0.07      & 0.6   & 4.9        &   % CO + Dust
 0.07      & 0.6   & 4.7        &   % CO + Dust + Ac. Rate
 0.07      & 0.4   & 3.7        &   % CO + Dust + Ac. Rate + Out limit
 0.07      & \textbf{0.5}   & 3.7        \\  % CO + Dust + Ac. Rate + Out limit + Strong AC.RATE
           &   & Age&                    &
   1.0     & 4.5   & 10.0       &   % CO + Dust
   1.0     & 5.5   & 10.0       &   % CO + Dust + Ac. Rate
   1.0     & 3.4   &  9.8       &   % CO + Dust + Ac. Rate + Out limit
   1.0     & \textbf{3.2}   & 8.6       \\  % CO + Dust + Ac. Rate + Out limit + Strong AC.RATE
           &   & $\Delta$Age    &     &
   0.05    &  4.8  & 9.0        &   % CO + Dust
   0.05    &  4.3  & 8.9        &   % CO + Dust + Ac. Rate
   0.05    &  2.4  & 8.8        &   % CO + Dust + Ac. Rate + Out limit
   0.05    &  \textbf{1.5}  & 5.4        \\  % CO + Dust + Ac. Rate + Out limit + Strong AC.RATE
           &   & M$_{disk}$     &     &
  0.03    &0.30   & 0.90       &
  0.03    &0.25   & 0.80       &
  0.03    &0.11   & 0.34       &
  0.03    &\textbf{0.10 } & 0.28    \\
           &   & N$_{models}$   &     &
        & 785 &         &   % CO + Dust
        & 563 &         &   % CO + Dust + Ac. Rate
        & 159 &         &   % CO + Dust + Ac. Rate + Out limit
        & \textbf{88}   &         \\  % CO + Dust + Ac. Rate + Out limit + Strong AC.RATE
\hline
  GM Aur  &   & $\beta$x$10^5$ &               &
  0.1     &  13   & 180      &   % CO + Dust
  0.1     &  4.8  &  25      &   % CO + Dust + Ac. Rate
  0.2     &  2.2  &  10      &   % CO + Dust + Ac. Rate + Out limit
  0.2     &   \textbf{1.8}   &   8       \\  % CO + Dust + Ac. Rate + Out limit + Strong AC.RATE
          &   & $j_{\cld}$&                &
 19.2     &  20.2 &  20.9      &   % CO + Dust
 19.3     &  20.2 &  20.9      &   % CO + Dust + Ac. Rate
 19.3     &  19.8 &  20.7      &   % CO + Dust + Ac. Rate + Out limit
 19.3     &  \textbf{19.8} &  20.3      \\  % CO + Dust + Ac. Rate + Out limit + Strong AC.RATE
           &   & $\rcrit$&               &
 4         &  920  & 4900       &   % CO + Dust
 5         &  930  & 4900       &    % CO + Dust + Ac. Rate
 5         &  105  & 1900       &   % CO + Dust + Ac. Rate + Out limit
 5         &  \textbf{85}   & 500       \\   % CO + Dust + Ac. Rate + Out limit + Strong AC.RATE
\hline
           &   & $\omega_{\cld}$&                &
 0.02      &  10   & 120        &   % CO + Dust
 0.02      &  10   & 120        &   % CO + Dust + Ac. Rate
 0.02      &  4.5  &  30        &   % CO + Dust + Ac. Rate + Out limit
 0.1       &   \textbf{4.3}   &  20        \\  % CO + Dust + Ac. Rate + Out limit + Strong AC.RATE
           &   & $T_{\cld}$&                &
 2         &  15   & 30         &   % CO + Dust
 2         &  16   & 30         &   % CO + Dust + Ac. Rate
 2         &  15   & 30         &   % CO + Dust + Ac. Rate + Out limit
 2         &  \textbf{15}   & 30         \\  % CO + Dust + Ac. Rate + Out limit + Strong AC.RATE
           &   & $M_{\cld}$&                &
 0.55      &  1.0  & 1.5       &   % CO + Dust
 0.55      &  1.0  & 1.5       &   % CO + Dust + Ac. Rate
 0.55      &  0.9  & 1.3       &   % CO + Dust + Ac. Rate + Out limit
 0.55      &  \textbf{0.9}  & 1.2       \\  % CO + Dust + Ac. Rate + Out limit + Strong AC.RATE
           &   & $J_{\cld}$&                &
 52.4      &  53.5 & 54.4      &   % CO + Dust
 52.5      &  53.5 & 54.4      &    % CO + Dust + Ac. Rate
 52.5      &  53.0 & 54.1      &   % CO + Dust + Ac. Rate + Out limit
 52.5      &  \textbf{52.9} & 53.5      \\  % CO + Dust + Ac. Rate + Out limit + Strong AC.RATE
           &   & AT&                    &
 0.08      &  0.7  & 5          &   % CO + Dust
 0.08      &  0.6  & 5          &   % CO + Dust + Ac. Rate
 0.08      &  0.6  & 4.8        &   % CO + Dust + Ac. Rate + Out limit
 0.08      &  \textbf{0.5}  & 4.3        \\  % CO + Dust + Ac. Rate + Out limit + Strong AC.RATE
           &   & Age&                    &
 1         &  4.4  & 10         &   % CO + Dust
 1         &  5.1  & 10         &   % CO + Dust + Ac. Rate
 1         &  4.5  & 10         &   % CO + Dust + Ac. Rate + Out limit
 1         &  \textbf{4.4}  & 10         \\  % CO + Dust + Ac. Rate + Out limit + Strong AC.RATE
           &   & $\Delta$Age     &     &
  0.05     &  4.9  & 8.9        &   % CO + Dust
  0.05     &  3.8  & 8.9        &   % CO + Dust + Ac. Rate
  0.05     &  1.6  & 6.1        &   % CO + Dust + Ac. Rate + Out limit
  0.05     &  \textbf{1.0}  &  3.9       \\  % CO + Dust + Ac. Rate + Out limit + Strong AC.RATE
           &   & M$_{disk}$     &     &
  0.023    &0.3    & 0.87       &
  0.023    &0.3    & 0.80       &
  0.023    &0.14   & 0.80       &
  0.028    &\textbf{0.12} & 0.35    \\
           &   & N$_{models}$   &     &
        &569&         &   % CO + Dust
        &426&         &   % CO + Dust + Ac. Rate
        &136&         &   % CO + Dust + Ac. Rate + Out limit
        &\textbf{ 77}&         \\  % CO + Dust + Ac. Rate + Out limit + Strong AC.RATE
\hline
\end{tabular}
\label{Table:Summary} \caption{Summary of the Monte-Carlo
exploration for different sets of constraints. The minimum,
maximum and mean value of the successful fitting models are given.
Symbols: AT Accretion time or molecular cloud collapse time.
Units: $j_{\cld}$ in cm$^2$s$^{-1}$, $\rcrit$ in AU,
$\omega_{\cld}$ in units of 10$^{-14}$ s$^{-1}$, $T_{\cld}$ in K,
$M_{\cld}$ and $M_{disk}$ in solar masses, $J_{\cld}$ in g
cm$^2$s$^{-1}$, and AT, Age and $\Delta$Age in Myrs.}
\end{table*}

Figure~\ref{ParameterSpace} constitutes a summary of our global
results for DM Tau and GM Aur for both parameterizations of
turbulence. Results are shown for three sets of observational
constraints (\{3\}, \{4\} and \{5\}). The most extended light-grey
contours correspond to the envelope in the $(\nu,j_{\cld})$ space
covered by models matching the CO and dust observations, and the
star accretion rate with the large error bar (set \{3\}). This
results in relatively weak constraints on the physical parameters.
Less restrictive observational constraints are therefore not shown:
For DM Tau $\alpha$ models, 30\% of all launched models satisfy the
CO data alone (set \{1\}) and 18\% satisfy the CO+dust data (set
\{2\}). If we include the accretion rate information we are able to
slightly reduce the number of fitting models to 15\% (set \{3\}).
These constraints are quite unrestrictive because they do not limit
the outer radius of the disk and unreasonably massive and extended
disks can result.
% TG: Inutile je pense
%\textbf{Most of these results are not well
%resolved since 1D models are not accurate for distances on the
%order of 1000 AU because $H/r\sim 0.35$ becomes too large.}

Adding a reasonable maximum value for the density of the disks at
their outer edge (set \{4\}) yields a significant reduction of the
permitted $(\nu,j_{\cld})$ space, either for the DM Tau $\beta$
model or for the GM Aur $\alpha$ and $\beta$ models. This is shown
as a dashed contour and grey color in fig.~\ref{ParameterSpace}. A
still narrower set of solutions can be achieved by restricting the
range of allowed accretion rates (set \{5\}). The final ensemble of
solutions is plotted as a thick colored contour in
fig.~\ref{ParameterSpace}. We consider that this set of solutions
provides the most realistic constraints on $\nu$ and $j_{\cld}$ for
DM Tau and GM Aur.

Figure~\ref{ParameterSpace} also shows that a right value of
$\alpha$, $\beta$ and $j_{\cld}$ is not a sufficient condition to
match the observations. The presence of other initial parameters
imply that only a fraction of the models really fit the imposed
conditions. This fraction peaks to 60\% for the central region in
the DM Tau $\alpha$ panel, 40\% for the $\beta$ case and 50\% and
30\% for GM Aur $\alpha$ and $\beta$ respectively.

The detailed values of the space of parameters allowed for sets
\{2\} to \{5\} are given on Table 5. We list the maximum, minimum
and statistic mean value of the model parameters that fit each set
of conditions.

Our first conclusion is that it is possible to use the estimated
current state of observed disks to get information about their
formation and evolution using relatively simple models. Globally
the constraints on the parameters are not very strong, but they
anyway provide useful information and ask for further detailed
observations of these objects. For instance, a reduction of the
star age uncertainty and/or the star accretion rate error bar
would be especially useful. Another important conclusion to be
drawn is that both $\alpha$ and $\beta$ models provide a
satisfactory evolutionary scenario for the disks around DM Tau and
GM Aur. This does not mean that either of these parameterizations
are correct but that none of them can be excluded on the basis of
these results.

The current disks around DM Tau and GM Aur can be obtained from very
different values of the angular momentum in the molecular cloud,
corresponding to values of the centrifugal radius spanning the
entire range (4 to 5000\,AU) allowed by our numerical approach. A
very small value of $\rcrit$ requires large values of the viscosity
to yield an efficient outward diffusion of the disk to match its
outer radius. A very large $\rcrit$ also requires a large viscosity,
but this time so as to yield a large enough accretion rate. As
expected, no useful constraint on $T_{\cld}$ can be derived.
Constraints on the initial rotation of the disk ($j_{\cld}$ or
$\omega_{\cld}$) are weak although models with relatively low values
of $j_{\cld}$ seem to be favored.

The mean derived value for the model parameters together with the
minimum and maximum allowed values are given on Table 5. The mean
value of $\alpha$ inferred from this study is 0.02 for DM Tau and
0.002 for GM Aur, both close to the \emph{standard} value of
$\alpha$=0.01 generally found in the literature (Hartmann et al,
1998). In the $\beta$ parameterization of turbulence the mean values
obtained are $9\times 10^{-5}$ and $2\times 10^{-5}$ for DM Tau and
GM Aur respectively. The $\beta$ value for GM Aur agrees with the
prediction that $\beta\approx 2 \times 10^{-5}$ from Richard and
Zahn (1999). The mean value of $\beta$ for DM Tau is a factor of
five larger but lower values close to the theoretical prediction are
allowed. These conclusions apply to set \{5\}, but remain valid for
sets \{2\} to {4\}.

A significant part of all models developed gravitational
instabilities near the end of the disk-formation era and mostly for
a relatively short time span of $1-2 \times 10^5$ yrs in systems
with ages less than $3 \times 10^5$ yrs. These thus have a limited
impact on the results. We also tested the importance of outflows in
the DM Tau $\alpha$ case by arbitarily imposing an inner boundary at
10 AU. The extra series of 500 models show no statistically
significant difference with the results presented here, except for
models with very low values of $j_{\rm cd}$.
% TG: En fait je ne suis pas sur d'avoir tout compris ce que tu as
% fait... A verifier donc.

% TG: Je pense que ce qui suit n'est pas vraiment utile (mieux vaut
% repondre au referee dans la lettre -l'article est deja assez long)
%Indeed we found models fitting our set {5} of constraints
%had centrifugal radius between 50 and 3600 AU with a mean
%statistical value of 650 AU.}

%\textbf{The disk masses for set \{6\} models vary by almost two
%orders of magnitude but most of the models have reasonable masses
%on the order of 0.008 M$_\odot$.}

%TG: commented out (discussed in next section)
%% Finally, the qualitative differences found in both systems from
%% using the $\alpha$ and $\beta$ parameterizations of turbulence
%% come mainly from the temperature dependance of viscosity on the
%% $\alpha$ models, contrary to the $\beta$ case. This makes the
%% $\alpha$ models strongly diffusive at initial phases when compared
%% with the $\beta$ cases.

\subsection{Inferred structures of DM Tau and GM Aur}

\begin{figure*}
\centering
\includegraphics[width=17cm]{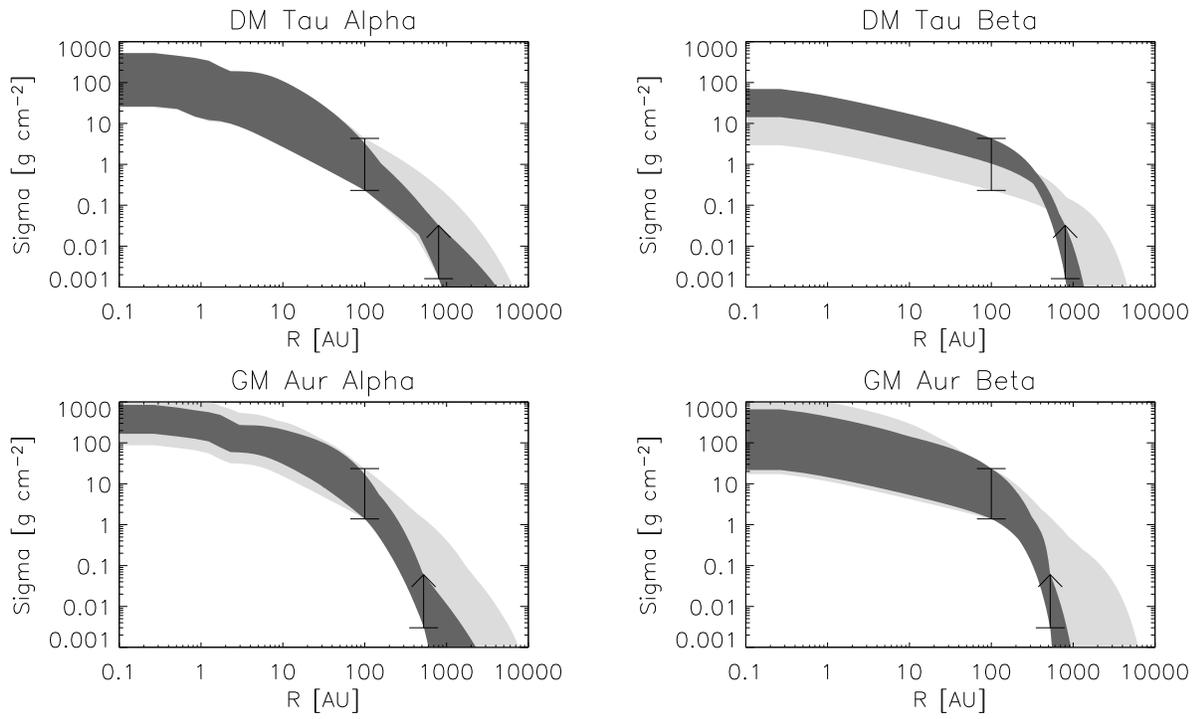}
\caption{Radial surface density profile of successful models of DM
Tau
  and GM Aur. The dark-shaded regions correspond to the models fitting
  the constraints \{5\}. The light-shaded regions represent the
  additional solutions when the constraints
  are only the CO + dust + Accretion rate with large error bar (set
  \{3\}). The error bars are the constraints on $\Sigma$ imposed by
  \{5\}.}
\label{ModelsSigma}
\end{figure*}

\begin{figure*}
\centering
\includegraphics[width=17cm]{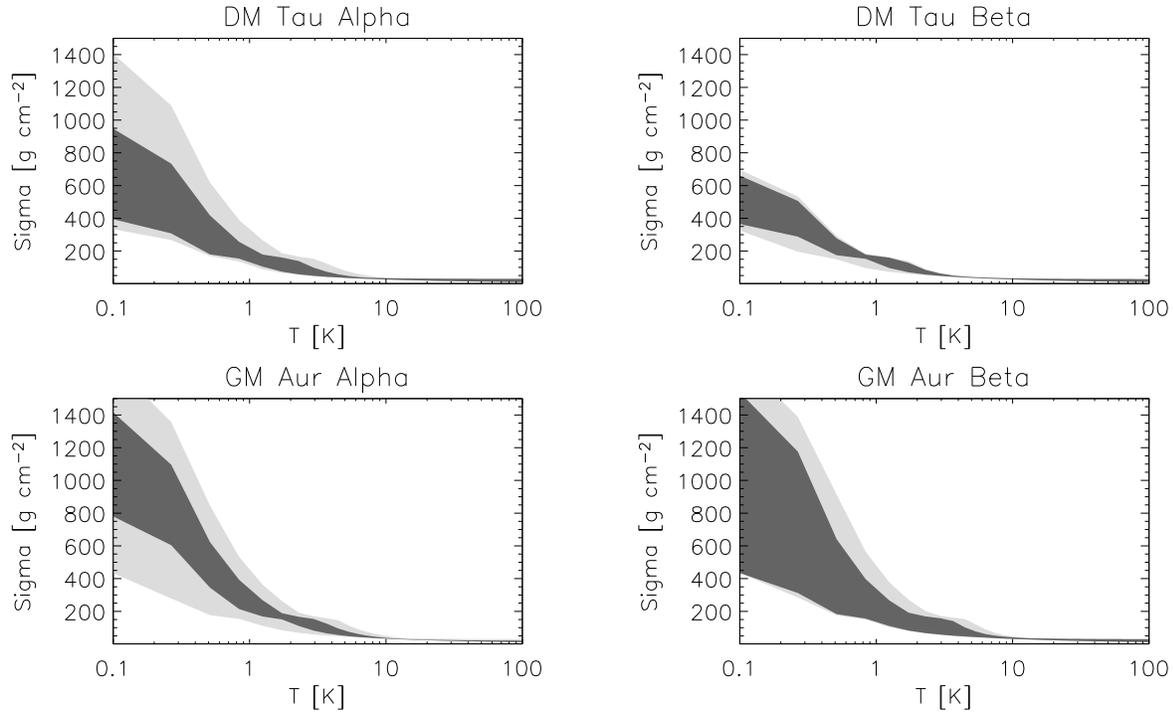}
\caption{Ensemble of possible mid-plane disk temperatures for two
different sets of observational constraints: \{3\}, light-shaded
and \{5\}, dark-shaded.} \label{ModelsTemp}
\end{figure*}

\begin{figure*}
\centering
\includegraphics[width=17.0cm]{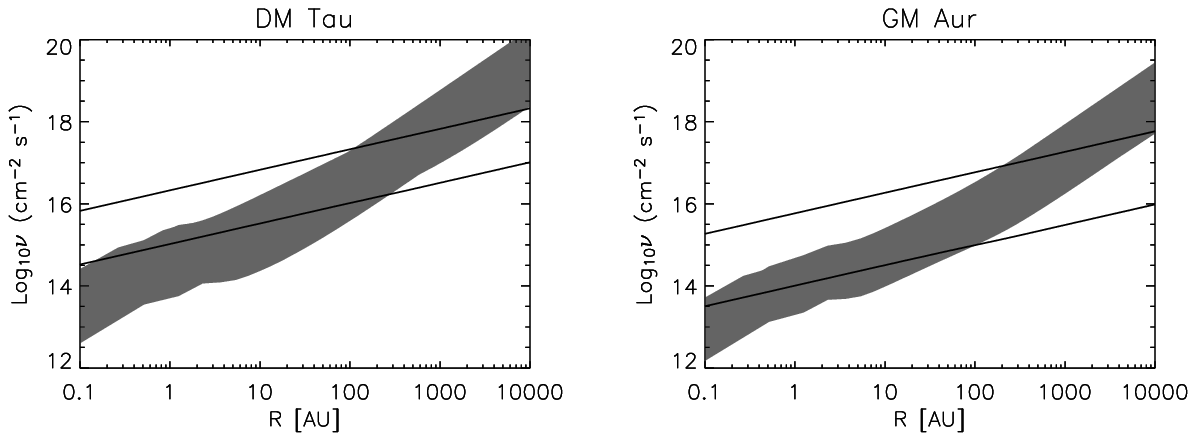}
 \caption{Viscosities of successful models of DM Tau and GM
 Aur. Shadowed regions correspond to models calculated with the
 $\alpha$ parameterization. Lines represent the upper and bottom limit
 of viscosities for the $\beta$ cases.}
 \label{FgViscosities}
\end{figure*}

The selected sets of models provide information not only on the
regions where we have observational constraints but also on the
general structure of the disks around DM Tau and GM Aur. Figure
~\ref{ModelsSigma} shows the totality of surface density distributions
for the four cases studied and different sets of constraints. The set
of models \{3\} and \{5\}, are represented as light- and superimposed
dark-shaded regions, respectively. Fitting the {\it absolute} outer
radii of the disks would provide an additional powerful
constraint. This may be difficult to obtain since dust observations
lack sensitivity and CO is affected by photodissociation and
condensation at low temperatures (Aikawa et al, 1999; Dartois et
al. 2003). Unfortunately, at present it is not clear what exactly the
outer radii inferred from the millimeter and optical observations
imply: they could either be interpreted as an abrupt or as a steady
decrease of the density profile. Without assuming a sharp outer edge
for GM Aur, models with unreasonable large disk masses become possible
fits to the observations (set \{3\}). On the other hand, sharp outer
edges of circumstellar disks would be in better agreement with the
$\beta$ prescription of turbulence, as shown by
fig.~\ref{ModelsSigma}.

Temperature is another important quantity whose evaluation
has important implications for planet formation processes. In the
$\alpha$ cases it also affects the intensity of the viscous
turbulence, $\nu$, and the disk evolution. The envelopes of all
possible radial profiles of temperatures for the fitted models are
shown in Figure~\ref{ModelsTemp}. These are the thermal profiles
of our calculated models at the time when they verify the selected
constraints. As in the previous figure, two constraints, \{3\} and
\{5\}, are shown as light- and dark-shaded regions, respectively.
At small radial distances, the temperature is mostly due to
viscous dissipation. At large radial distances, the disk becomes
optically thin and the temperatures are essentially controlled by
the stellar flux.
%TG: Si la suite est vraie, il faudrait le dire dans la section
%``temperature''. Pas besoin d'en parler ici.
%%  and a fixed background temperature taken as the initial
%% temperature of the molecular cloud $T_{\cld}$.
The $\beta$ models are
significatively cooler than the $\alpha$ models because they have less
mass in the inner disk, implying lower optical depth and lower central
temperatures for a given effective (photospheric) temperature. For the
same reasons, GM Aur appears to be warmer than DM Tau.
% Ca n'est pas necessaire, meme si on le fera effectivement
%% The history evolution of the
%% thermal properties of the disk is of great importance to the
%% evaluation of the production of planetesimals in these systems and a
%% detailed analysis will be presented elsewhere (Hueso and Guillot, work
%% in preparation).

Turbulent viscosity is the key variable to understand the disk
evolution. Figure~\ref{FgViscosities} shows the envelope of
possible $\nu$ profiles as a function of stellar distances for the
solutions that fit the observations of DM Tau and GM Aur (set
\{5\}). At first glance, the solutions for the $\alpha$ and
$\beta$ models may appear to be relatively similar, a consequence
of the constraint on the density profile provided by the
observations. However, marked departures between the two profiles
in the $[10-1000\rm\,AU]$ region are apparent. Globally,
$\nu_{\alpha}$ varies as $r^{3/4}$, while $\nu_{\beta}$ varies as
$r^{1/2}$, meaning than the $\alpha$ models are diffused more
efficiently in the outer disk. This explains why, at large
distances from the star, the density profile decreases less
sharply in the $\alpha$ models than in the $\beta$ models.

\subsection{Global evolution of the disk and gravitational instabilities}

\begin{figure*}
\centering
\includegraphics[width=17cm]{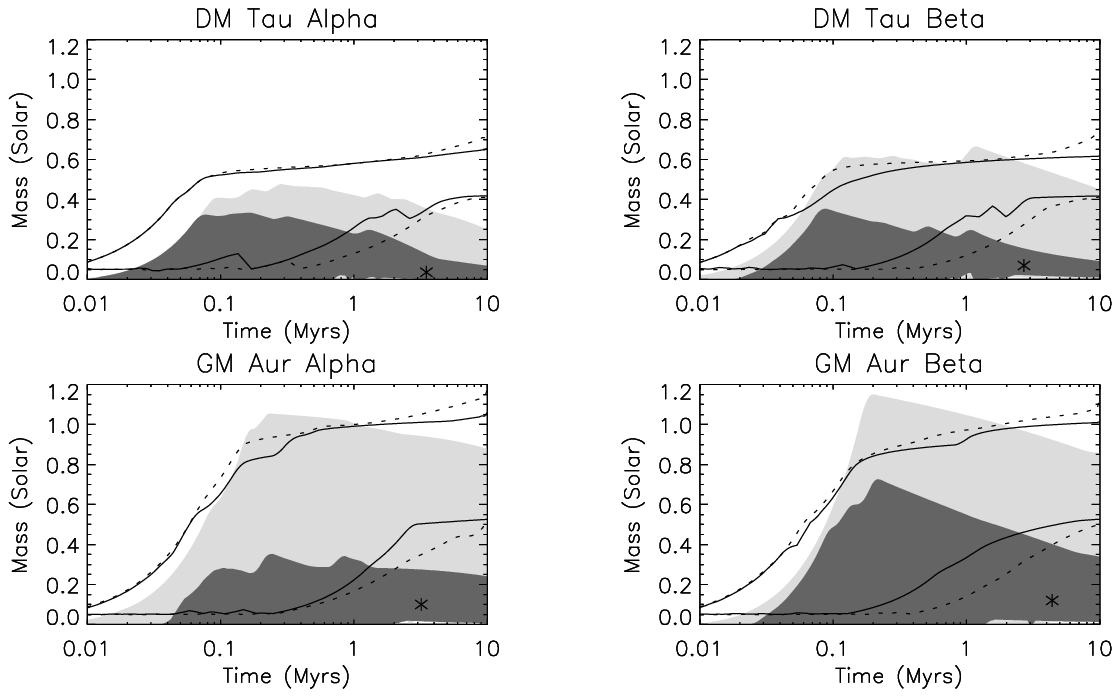}
\caption{Star mass and disk mass as a function of time for models
succefully fitting selected observational contraints. The light- and
dark-shaded regions show the evolution of the disk mass with
constraints \{3\} and \{5\}, respectively. The dotted and plain
lines represent the envelopes of possible star masses with the same
constraints. The star symbols mark the mean values of the ages and
disk masses of models fulfilling set \{5\}.} \label{FgStardiskmass}
\end{figure*}

Figure~\ref{FgStardiskmass} shows the envelope of possible star
and disk mass evolution for the models matching conditions \{3\}
and \{5\}, respectively.  Up to 15\% of our selected models (set
\{5\}) were gravitationally unstable, though generally only over a
short fraction of their evolution, near the end of the disk
formation era. These correspond to the models for which the mass
of the disk is close to or exceeds the central star mass. The
gravitational instability arises at 30$\pm$10 AU and times
$1.0\pm0.6\times 10^5$\,yr and typically lasts for $10^4-10^5$ yrs
for DM Tau and less than $2\times 10^5$ for GM Aur due to the
larger amount of mass the disk has to process.
%% We expect that gravitational instabilities would
%% produce stronger dissipative effects on this early period than
%% considered in the model but, eventually, these disks return to
%% gravitational stable situations were they would evolve viscously
%% erasing most of the structure developed on this early phase.
Because we compare observations and models at ages larger than
$10^6$ yrs, gravitational instabilities should not
affect signficantly our analysis. In sets \{4\} and \{5\}, models
with $\alpha > 0.03$ were never gravitationally unstable. Models
with lower values of $\alpha$ had an imposed $\alpha=0.03$ during
gravitational instable periods (see section~3.6.3).

% TG: (IMPORTANT)
% Pour voir si effectivement les instabilites gravitationnelles n'ont
% pas d'effet, il faudrait montrer que les modeles alpha avec
% instabilites gravitationnelle sont statistiquement similaires aux
% modeles sans instabilite gravitationnelle. Ou mieux, comparer les
% resultats avec et sans l'augmentation artificielle de alpha.

%% We did not
%% consider any such treatment of the gravitationally unstable cases in
%% the $\beta$ models.  Still, we found the same error bar in the
%% determination of $\beta$. This tells us that, with the given
%% uncertainty in observational data, gravitationally unstable processes
%% in DM Tau and GM Aur did not have a great importance in their global
%% evolution or, at least, that these effects cannot be resolved from
%% their current structure.

% TG: A VERIFIER !!!
The problem of the evolution of the mass of accretion disks as
  a function of viscosity and initial angular momentum $J_{\cld}$ was
  also studied by Nakamoto and Nakagawa (1995). These authors derived
  a criterion on the viscosity parameter $\alpha$ which guarantees
  that at any given time, the disk to star mass ratio never exceeds
  0.1:
\begin{equation}
\log \alpha \geq -2.0 + 4.5 \log J_{\cld} -3.9(\log J_{\cld})^2,
\end{equation}
with $J_{\cld}$ in units of 10$^{53}$ g cm$^2$ s$^{-1}$. Their
study is relevant for systems of $\sim 1\rm\,M_{\odot}$ collapsing
in $\sim 6 \times 10^5$ yrs. They considered the parameter range
$J_{cd}=[6\times 10^{52}-10^{53}]$\,g cm$^2$\,s$^{-1}$ and
$\alpha=[10^{-5}-0.1]$.

For our simulations of DM Tau $\alpha$ we plot on
Figure~\ref{FgNakamoto}, the regions of the parameter space that
have lower masses than 0.1 M$_{*}$ at the end of the disk formation
period (and hence, at the moment of maximum disk mass). Our results
compare well with those of Nakamoto and Nakagawa, though their
result is strictly valid for systems collapsing in 0.6 Myr years and
ours extend to systems forming in 0.05 to 5 Myr. A comparison with
fig.~\ref{ParameterSpace} shows that most solutions for DM Tau and
GM Aur are to the left of the critical line, i.e. the disk to star
mass ratio has been, or in some cases is still, larger than $0.1$.
This seems to contradict the fact that most observed disks appear to
have values of this ratio smaller than 0.1 (e.g. Beckwith et al.
1990; Kitamura et al. 2002). However, this contradiction may only be
apparent, given the fact that the dust absorption coefficient at
millimeter wavelengths may have been overestimated (see
section~\ref{sec:constraints}), and/or that the time spent with a
large disk mass is generally relatively short. It is also possible
that DM Tau and GM Aur are, among protoplanetary disks, exceptions,
and that most other disks have suffered more from either tidal
interactions with nearby stars and/or photoevaporation from
external, relatively massive stars. Alternatively, our assumptions
regarding the uncertain opacity factor may be too conservative and
allow for unrealistically large values of the disk mass. This would
imply that stronger constraints can be derived from the observations
presently available.

\begin{figure}
\centering
\includegraphics[width=8.5cm]{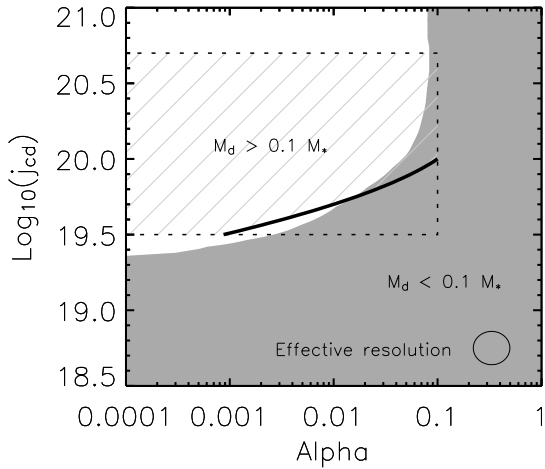}
\caption{Regions of the parameter space where massive disk and
non-massive disks can be found. The constraint given by Nakamoto
and Nakagawa traces a region (dashed) where massive disks at the
end of the molecular cloud collapse are to be found. The dotted
box is the area of the space of parameters covered by their study.
The grey shaded area represents the region where we found models
with disks less massive than 0.1 $M_{*}$ at the end of the
collapse period. The spatial resolution of our calculations is
shown by the lower right circle.} \label{FgNakamoto}
\end{figure}

\subsection{Results for the collapse of magnetized molecular clouds}

In order to assess the importance of the collapse model in the
final results, we also study an $\alpha$ model of DM Tau in which
the molecular cloud is assumed to collapse under the influence of
a strong magnetic field. Figure~\ref{FgBasu} shows the final
characteristics of models fitting the observational constraints
when eqs.~(\ref{eq:mdot-basu},\ref{eq:rd-basu}) are used.
Table~\ref{TbBasu} shows a summary of our results when fitting the
observational constraints \{2\} to \{5\}.

\begin{figure}
\centering
\includegraphics[width=8.5cm]{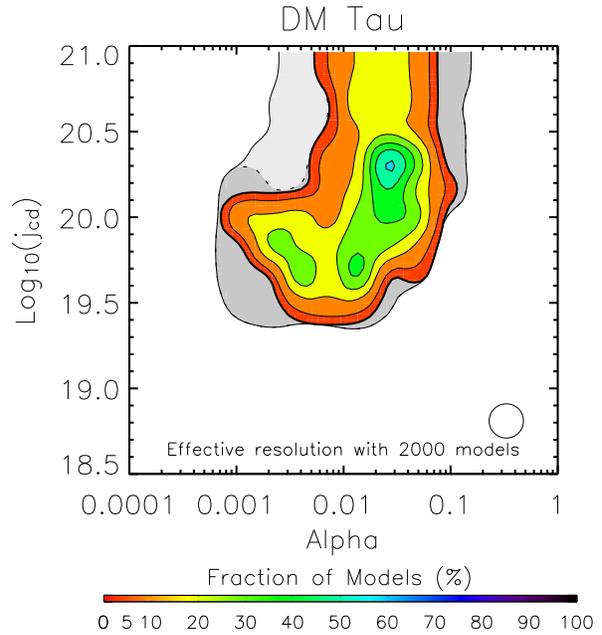}
\caption{Distributions of models fitting the sets of constraints
\{3\}, \{4\} and \{5\} for DM Tau $\alpha$ in the case of an
initially magnetized molecular cloud. See fig.~11 for details.}
\label{FgBasu}
\end{figure}

\begin{figure}
\centering
\includegraphics[width=8.5cm]{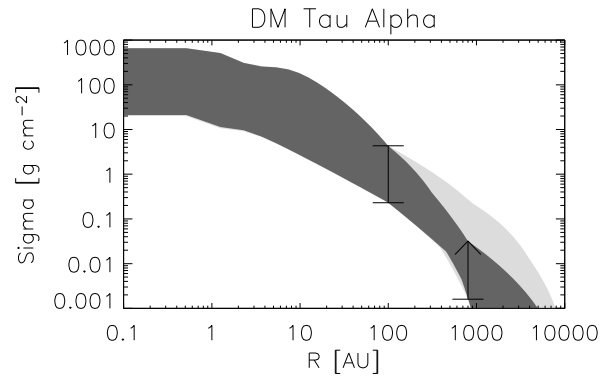}
\caption{$\Sigma$ surface density distributions at different times
for the successful models in the magnetic cloud case, fulfilling the
constraints \{3\} (light-grey) or \{5\} (dark-grey). See
fig.~12 for details.} \label{BasuSigma}
\end{figure}

\begin{figure}
\centering
\includegraphics[width=8.5cm]{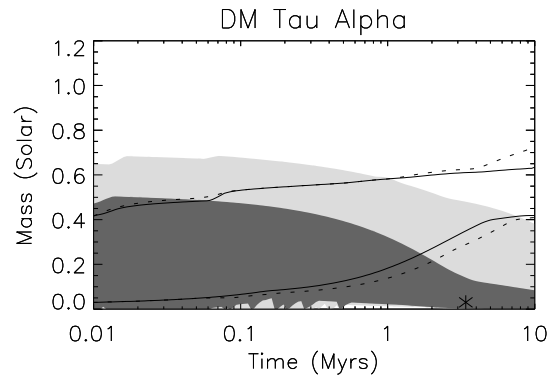}
\caption{Star and disk mass evolution for the successful models in
the magnetic cloud case. See fig.~15 for details.}
\label{BasuStardiskmass}
\end{figure}

\begin{table*}
\begin{tabular}{p{0.5cm}p{0.40cm}p{0.5cm}p{0.4cm}p{13.7cm}}
\hline \vspace{0.005in}\textbf{System}     &
 &
\vspace{0.005in}\textbf{Param.}  &
 &
\vspace{0.005in}\textbf{Model results obtained for the different observational constraints} \smallskip \\
\hline
\end{tabular}

\begin{tabular}{p{0.7cm}p{0.1cm}p{0.6cm}p{0.4cm}
                p{2.9cm}p{2.9cm}p{3.4cm}p{3.7cm}}
 &   &   &  & \textbf{\{2\}:} CO + Dust
            & \textbf{\{3\}:} \{2\} + $\dot{M}$
            & \textbf{\{4\}:} \{3\} + Outer Limit
            & \textbf{\{5\}:} \{4\} + Accurate $\dot{M}$   \\
\end{tabular}

\begin{tabular}{p{1.20cm}p{0.0cm}p{0.45cm}p{0.0cm}
                p{0.7cm}p{0.8cm}p{0.8cm}    %  CO + Dust
                p{0.7cm}p{0.8cm}p{0.8cm}    %  CO + Dust + Ac. Rate
                p{0.7cm}p{0.8cm}p{0.8cm}    %  CO + Dust + Ac. Rate + Out limit
                p{0.7cm}p{0.8cm}p{0.8cm}}   %  CO + Dust + Ac. Rate + Out limit + Strong AC.RATE
 &   &   &           &
    Min &  Mean &  Max &
    Min &  Mean &  Max &
    Min &  Mean &  Max &
    Min &  \textbf{Mean} &  Max \\
\hline
  DM Tau    &   & $\alpha$      &               &
7x$10^{-4}$ & 0.03  & 0.2       &   % CO + Dust
7x$10^{-4}$ & 0.02  & 0.1       &   % CO + Dust + Ac. Rate
8x$10^{-4}$ & 0.03  & 0.1       &   % CO + Dust + Ac. Rate + Out limit
    0.004   & \textbf{0.02}  & 0.1       \\  % CO + Dust + Ac. Rate + Out limit + Strong AC.RATE
        &   & $j_{\cld}$&                &
  19.6    &20.5   & 21.2       &   % CO + Dust
  19.6    &20.5   & 21.2       &   % CO + Dust + Ac. Rate
  19.6    &20.3   & 21.1       &   % CO + Dust + Ac. Rate + Out limit
  19.6    &\textbf{20.3}  & 21.1   \\  % CO + Dust + Ac. Rate + Out limit + Strong AC.RATE
           &   & $\rcrit$&      &
  5.4      &80         & 240   &   % CO + Dust
  5.4      &70         & 240   &   % CO + Dust + Ac. Rate
  5.4      &55         & 200   &   % CO + Dust + Ac. Rate + Out limit
  5.4      &\textbf{50}& 200   \\   % CO + Dust + Ac. Rate + Out limit + Strong AC.RATE
\hline
           &   & $\omega_{\cld}$&                &
    0.7    &  9    & 26       &   % CO + Dust
    0.7    &  8    & 26       &   % CO + Dust + Ac. Rate
    0.7    &  7    & 26       &   % CO + Dust + Ac. Rate + Out limit
    0.7   &  \textbf{6}   & 25    \\  % CO + Dust + Ac. Rate + Out limit + Strong AC.RATE
           &   & $T_{\cld}$&                &
     2     &  14 & 30         &   % CO + Dust
     2     &  13 & 30         &   % CO + Dust + Ac. Rate
     2     &  15 & 30         &   % CO + Dust + Ac. Rate + Out limit
     2     &  \textbf{15} & 30     \\  % CO + Dust + Ac. Rate + Out limit + Strong AC.RATE
           &   & $M_{\cld}$&                &
  0.41     & 0.60 & 0.97      &   % CO + Dust
  0.41     & 0.58 & 0.86      &   % CO + Dust + Ac. Rate
  0.41     & 0.54 & 0.67      &   % CO + Dust + Ac. Rate + Out limit
  0.42     & \textbf{0.54} & 0.67      \\  % CO + Dust + Ac. Rate + Out limit + Strong AC.RATE
           &   & $J_{\cld}$&                &
  52.5     &53.6   & 54.5   &   % CO + Dust
  52.5     &53.5   & 54.4   &   % CO + Dust + Ac. Rate
  52.5     &53.4   & 54.2   &   % CO + Dust + Ac. Rate + Out limit
  52.5     &\textbf{53.4}   & 54.2      \\  % CO + Dust + Ac. Rate + Out limit + Strong AC.RATE
           &   & AT&        &
  0.05     &0.9    & 9.5    &   % CO + Dust
  0.05     &0.8    & 9.3    &   % CO + Dust + Ac. Rate
  0.05     &0.5    & 9.3    &   % CO + Dust + Ac. Rate + Out limit
  0.05     &\textbf{0.5}    & 9.3       \\  % CO + Dust + Ac. Rate + Out limit + Strong AC.RATE
           &   & Age&                    &
  1.5      &3.8    & 7.1    &   % CO + Dust
  1.5      &4.0    & 7.1    &   % CO + Dust + Ac. Rate
  1.5      &3.5    & 7.0    &   % CO + Dust + Ac. Rate + Out limit
  1.5      &\textbf{3.4}    & 7.0       \\  % CO + Dust + Ac. Rate + Out limit + Strong AC.RATE
           &   & $\Delta$Age     &     &
  0.05     &2.3    & 5.6    &   % CO + Dust
  0.05     &2.1    & 5.6    &   % CO + Dust + Ac. Rate
  0.05     &1.7    & 5.5    &   % CO + Dust + Ac. Rate + Out limit
  0.05     &\textbf{1.0}    & 2.7        \\  % CO + Dust + Ac. Rate + Out limit + Strong AC.RATE
           &   & M$_{disk}$     &     &
  0.002    &0.10   & 0.44       &
  0.008    &0.08   & 0.44       &
  0.008    &0.04   & 0.11       &
  0.01     &\textbf{0.03} & 0.11    \\

           &   & N$_{models}$   &     &
       &332    &         &   % CO + Dust
       &277    &         &   % CO + Dust + Ac. Rate
       &184    &         &   % CO + Dust + Ac. Rate + Out limit
       &\textbf{129}     &      \\  % CO + Dust + Ac. Rate + Out limit + Strong AC.RATE
\hline
\end{tabular}
\caption{Summary of Monte-Carlo exploration in the case of an initial
  collapse of a magnetized molecular cloud core. For each set of
constraints described in the text \{2\}, \{3\} \{4\} and \{5\} the
minimum, maximum and mean value of the successful fitting models
are given. Symbols: AT Accretion time or molecular cloud collapse
time. Unities: $j_{\cld}$ in cm$^2$s$^{-1}$, $\rcrit$ in AU,
$\omega_{\cld}$ in units of 10$^{-14}$ s$^{-1}$, $T_{\cld}$ in K,
$M_{\cld}$ and $M_{disk}$ in solar masses, $J_{\cld}$ in g
cm$^2$s$^{-1}$, and, AT, Age and $\Delta$Age in Myrs.}
\label{TbBasu}
\end{table*}

Compared with the hydrodynamical collapse models, these are
characterized by lower values of the initial rotational velocity
($6\times10^{-14}$ s$^{-1}$ vs.  $23\times10^{-14}$ s$^{-1}$) and
similar values of the specific angular momentum (Log$_{10}
(j_{\cld})=$ 20.3 vs. 20.0). This is because differential rotation
yields a faster rotation of the inner molecular cloud core.
Furthermore, the final centrifugal radii of the disks are smaller
than in the non-magnetic case for a given rotational velocity. The
only possible scenario for the formation of DM Tau is an inner
disk of intermediate size ($\rcrit\sim$ 50 AU) that expands in
time while it diffuses. This is true even if one considers only
the weak constraints \{2\}.

The distribution of surface densities for the successfully fitting
models is shown on fig.~\ref{BasuSigma}. Appart from the fact that
high values of the initial angular moment in the cloud core are
allowed, the results are similar to those obtained for the
non-magnetic case (fig.~\ref{ModelsTemp}), confirming that a strong,
sustained viscous evolution of the system is required to fit the
observations. In particular, the values of $\alpha$ derived are
almost identical regardless of the collapse scenario.

The global evolution of the star and disk masses is
shown on fig.~\ref{BasuStardiskmass}. Compared to the non-magnetic case
(fig.\ref{FgStardiskmass}), the 10 times faster accretion implies that
the disks grow very rapidly and are likely to become gravitationally
unstable early on.

%% RH: Not relevant or it could be better summarized
%% About half of these successful models were gravitationally
%% unstable at times relatively low ($2\times10^4$ yrs) and large
%% distances (300 AU) during less than $1\times 10^5$ yrs.

% TG: Ca n'apporte pas grand chose
%We expect similar results for a corresponding analysis of GM Aur,
%i.e. the solutions coming from a differential rotation molecular
%cloud would tend to stronger support the scenario of inner disks
%of intermediate size expanding to larger disks of reduced external
%mass.

\subsection{Turbulence driven by vertical convection?}

One of the advantages of our Monte-Carlo approach is that we can
test if some of the different theoretical mechanisms able to
produce turbulence in the disk can explain the observations of
extended disks. Here we test if the simplest characteristics of
thermal convection (Lin and Papaloizou, 1980) can explain the
current state of DM Tau. If thermal convection is the dominant
source of turbulence in circumstellar disks, the instability
should be essentially limited to the inner optically thick disk
with a more or less sharp transition in the outer optically thin
disk (Ruden and Pollack, 1991).

It is useful to consider some order of magnitude figures. In the
outer region the opacities are dominated by ice grains. For typical
temperatures $T\sim 10$\,K, the Rosseland opacity (gas+dust) is
$\kappa_{R}\sim 0.02\rm\,cm^2\,g^{-1}$ (Pollack et al. 1986). The
disk is unable to sustain a convective vertical transport of heat at
an optical depth $\tau \wig{<} 1$ implying a critical surface
density $\Sigma_{crit}\sim 2/\kappa_{R}\sim 100\rm\,g\,cm^{-2}$. A
circumstellar disk of mass 0.1 M$_{\odot}$ could then be convective
up to a maximum limit of $\sim50$\,AU. If we consider higher
temperatures and disk masses ($T=30$ K and
$M_{d}=0.3\rm\,M_{\odot}$) this limit extends to 280 AU. Therefore,
it would be extremely difficult to produce the extended structures
of disks like DM Tau and GM Aur by the outward diffusion of
initially smaller disks. If turbulent convection is the dominant
source of viscosity, these systems should have been formed
essentially as they are observed now.

In order to test this possibility we launched a new set of
calculations for DM Tau following the same range of values as in
the previous cases and considering an $\alpha$ constant in the
inner disk but decreasing in the outer part of the disk as a function
of the optical depth to a given power:
\begin{equation}
\alpha=\alpha_{0}\left(\frac{\tau}{\tau_{\rm crit}}\right)^n; \qquad
\tau\le \tau_{\rm crit}.
\label{eq:tau_crit}
\end{equation}
Here $\alpha_{0}$ characterizes the viscosity in the optically
thick disk, $n$ is an integer number that measures the transition
of the turbulent active to non active region and $\tau_{crit}=1.8$
following Ruden and Pollack (1991). We launched 900 models with
$n=8$ (sharp transition) and 500 with $n=1$ (smooth transition)
for DM Tau.

In the first case, $n=8$, we obtained 4 models able to
simultaneously satisfy the CO and dust constrains (set \{2\}) with
centrifugal radius around 1300 AU. These models were typically far
too massive and could not satisfy simultaneously the observed
accretion rate.

In the second case, we chose $n=1$. This situation corresponds to
a slow decrease of turbulence proportional to the optical depth.
It is probably unrealistic but could possibly model a hybrid case
in which convection is active at $\tau\wig{>}1$ and another
undefined, weaker, turbulent source sets in at lower optical
depths. In this case, material can be transported from the
optically thin to the optically thick regions.

% RH: New observational constraints changed this paragraph
%
%%We found that a reduced number of 7 models are able to fit the CO
%%+ dust requirements (set \{2\}). These models are characterized by
%%values of $\alpha=[0.04,0.25]$, $Log_{10}(j_{\cld})=[19.7-20.0]$
%%cm$^2$ s$^{-1}$ and $\rcrit=[50,240]$ AU. Only one of these models
%%is also able to satisfy the tight accretion rate and outer disk
%%edge assumed in our final set of conditions (set \{5\}). The
%%global evolution of this model together with the radial
%%distributions of its surface density and the disk effective
%%viscosity are shown in figs.~20, 21 and 22.

We found that 19 \% of all models are able to fit the CO + dust
requirements (set \{2\}) and that 13 \% also fit the accretion rate
constraint (set \{3\}). This is similar to the results obtained
previously with a uniform $\alpha$ value. However, the models
fitting the constraints \{3\} are now characterized by larger values
of the centrifugal radius ($\rcrit=[45,4900]$ AU), specific angular
momentum ($Log_{10}(j_{\cld})=[19.7-20.8]$) and hence viscosity
($\alpha=[0.002,0.6]$) than those listed in Table 5. These disks are
generally massive, they develop gravitational instabilities and are
formed basically as they are observed. We found that only a tiny
fraction $\sim 2$\% of all models also fit constraints \{4\}. These
are characterized by: $\rcrit=[45,300]$ AU; $\alpha=[0.003,0.07]$;
$Log_{10}(j_{\cld})=[19.7-20.1]$.  Finally, only one single model
was able to also fit simultaneously the strict accretion rate we
have considered on set \{5\}.

\begin{figure}
\includegraphics[width=8.0cm]{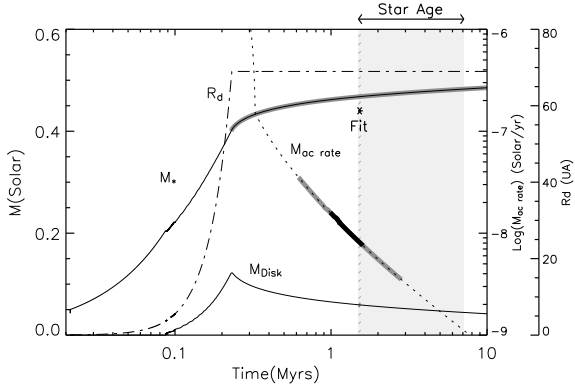}
\label{Figturbulencea} \caption{Global evolution of the star and
disk mass. This was the only case found where turbulence of
convective origin was found to satisfy the most retrictive set of
constraints \{5\}. See fig.~5 for details.}
\end{figure}

\begin{figure}
\includegraphics[width=8.5cm]{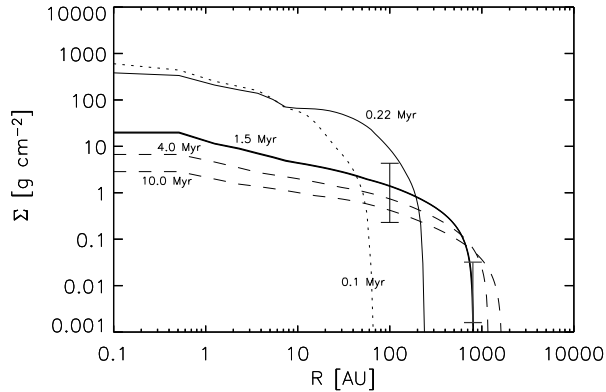}
\label{Figturbulenceb} \caption{$\Sigma$ surface density
distributions at different times for the same model as the
previous figure. See fig.~6 for details.}
\end{figure}

\begin{figure}
\includegraphics[width=8.5cm]{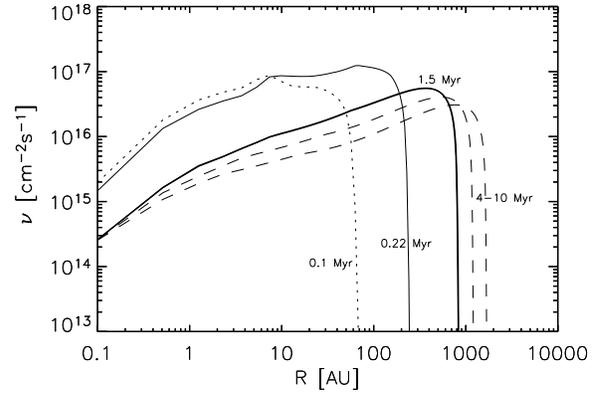}
\label{Figturbulencec} \caption{Radial distribution of turbulent
viscosity in the disk as a function of time. As the disk spreads
it is able to develop viscosity even in regions far beyond the
final centrifugal radius $\rcrit$=62 AU. This is because this case
corresponds to $n=1$ (see. eq.~(\ref{eq:tau_crit})), implying that
the transition between the convective to non-convective regime is
extremely smooth (and probably unrealistically so).}
\end{figure}

The global evolution of this model together with the radial
distributions of its surface density and the disk effective
viscosity are shown in figs.~20, 21 and 22. The model is
characterized by $\alpha=0.07$, $j_{\cld}=19.73$, $\rcrit=62$ AU
and total accretion time 0.22 Myr.  The molecular cloud parameters
are also found to agree reasonably well with those inferred for
the Taurus formation region ($\omega_{\cld}=4.38\times 10^{-14}$
s$^{-1}$, $T_{\cld}=12$ K, $M_{0}=0.05$ M$_{\odot}$ and
$M_{\cld}=0.527$ M$_{\odot}$). It must be stressed however that
this chance result requires initial conditions that are very close
to those listed here.

We conclude that it is extremely unlikely that convective
instabilities could have been the dominant source of angular
momentum transport in DM Tau and GM Aur.

\section{Conclusions}

We presented a (relatively) simple model of evolution of
circumstellar disks and applied this model systematically to two
well-characterized objects: DM Tau and GM Aur. We chose to survey
extensively a rather large parameter space, using various
observational constraints from CO observations of these disks,
dust emission properties, derived accretion rates and age and
masses of the central stars. Two viscosity parameterizations,
$\alpha$ and $\beta$, and two cloud collapse models were used.

We showed that the current state of DM Tau and GM Aur can be
understood as resulting from the collapse of a molecular cloud core,
the formation of an accretion disk and its spreading due to angular
momentum conservation and turbulent diffusive transport. This scenario
is consistent with a specific angular momentum initially in the cloud
core ($j_{\cld}\sim 10^{20}-10^{21}\rm\,cm^2\,s^{-1}$) similar to
measurements in class I protostellar cores (Ohashi et al.
\cite{Ohashi97}), and values of the viscosity that are within an order
of magnitude of the standard $\alpha\sim 0.01$ derived from measured
ages and accretion rates (e.g. Hartmann et al.
\cite{Hartmann98}). The significant outward diffusion implies that
most of the material has been reprocessed and lies far from the
position where it had originally fallen onto the disk.

More specifically, the values of $\alpha$ that can be inferred from
this study range within $0.001-0.1$ for DM Tau and
$4\times10^{-4}-0.01$ for GM Aur. In the case of the $\beta$ models,
the likely values of this parameter are $2 \times 10^{-5}-5 \times
10^{-4}$ for DM Tau and $2 \times 10^{-6}-8\times 10^{-5}$ for GM
Aur.

Unfortunately, the still relatively large error bars on
observationally determined quantities (in particular the age of the
system and the star accretion rate) prevent inferring tight
constraints on parameters characterizing the systems. We cannot argue
in favor of either the $\alpha$ or $\beta$ parameterizations of
turbulent viscosity. Similarly, our results are consistent with both a
relatively slow (hydrodynamic) or a fast (magnetic) collapse of the
primordial molecular cloud core. We can however rule
out thermal convection as the main source of turbulence in the
disk: The observed disks of DM Tau and GM Aur are too large and extend
too far in the optically thin regime in which convection would be
suppressed. On the other hand, they cannot be formed directly from the
molecular cloud. Convection could still play a role in the inner disk,
but another source of turbulence is required in the optically thin
regime.

%%Imposing a lower limit of $j_{\cld}\wig{>}
%%10^{20}\rm\,cm^2\,s^{-1}$, as implied by Ohashi et al.
%%(\cite{Ohashi97}) yields unchanged constraints on $\alpha$ for DM
%%Tau, but $0.004 < \alpha < 0.04$ for GM Aur. In the case of the
%%$\beta$ parameterization, it implies $4\times 10^{-4}<\beta
%%<2\times 10^{-4}$ for DM Tau, and a very limited set of solutions
%%with $\beta\sim 10^{-5}$ for GM Aur. }

The current state of both systems could be explained by a wide
variety of evolutionary histories. In most cases, the disks were
always gravitationally stable. However, for low- and high-values
of the initial angular momentum $j_{\cld}$ (corresponding to
$\rcrit\wig{<}10\,$AU or $\rcrit\wig{>}2000\,$AU), disks can
become gravitationally unstable for $\sim 10^5$\,yrs or less. In
the case of DM Tau and GM Aur, gravitational instabilities thus
appear to have had a limited role.

The relatively large values of the viscosity that we derive both for
DM Tau and GM Aur imply that an efficient turbulent diffusion
mechanism is present throughout these disks. As a consequence, any
giant planet forming in these disks will be in risk of migrating
rapidly into its star, except if (i) it is protected by an inner dead
zone characterized by a low viscosity or (ii) it forms late, i.e. when
the local disk mass is comparable to the mass of the planet itself. We
regard the second possibility as more likely, but this requires more
specific studies. In any case, the presence of extended viscous disks
around DM Tau and GM Aur has consequences on how we should apprehend
planet formation and should be included in future studies of this
problem.

Several improvements to this work are possible. Obviously, our
sets of constraints \{1\} to \{5\} should eventually be replaced
by a global assessment of the compatibility of the derived
density profiles with the different observations (SED, dust
emissivity, CO observations, reflected light observations). This
is not trivial because of possible variations of the dust to gas
ratio (dust migration due to gas drag, condensation at different
temperatures), of radial and vertical variations of the size
distribution of solid particles, and of variations in the chemical
composition (in particular due to CO condensation and its
photodissociation). Another improvement would be to include the
thermal evaporation of the disks at large orbital distances
(hundreds of AU).

In any case, we believe that meaningful constraints on
  physical processes leading to the formation of extended disks can
  now be derived. These critically depend on the inferred surface
  density profiles at 100 AU and beyond. High sensitivity observations
  able to detect small amounts of gas or dust in the outer regions of
  the disks such as forseen for ALMA are of course highly desirable. A
  better understanding of the inner disk regions and constraints on
  the rate of accretion onto the star would also be very valuable.
  Ongoing programs (in particular observations with the Spitzer
  telescope, but also from ground-based facilities) should ensure a
  rapid progress towards understanding the formation and evolution of
  protoplanetary disks.

%Future projets should provide observations that are directly
%relevant to the characterization of circumstellar disks and their
%evolution. The next generation of millimeter-wave interferometers,
%such as ALMA (the Atacama Large Millimeter Array) will be able to
%resolve details on much smaller scales on nearby accretion disks.
%The VLTI (Very Large Telescope Interferometer) will also
%contribute to this aim in the infrared. These observations with a
%high spatial resolution will allow studying the presently
%inaccessible inner regions of the disks. Furthermore, observations
%from monolithic telescopes in the infrared (Spitzer, Herschel,
%SOFIA), and with high-sensitivity coronographic devices
%(VLT-Planet Finder) should provide new information on the
%structure of these disks.

\begin{acknowledgements}
We thank P. Cassen for his comments and suggestions when we took
on this project, and A. Dutrey and S. Guilloteau, for providing us
with data in advance of publication. The manuscript also benefited
from constructive and useful comments from an anonymous referee.
We are grateful for discussions with P. Andr\'e, P. Hennebelle, B.
Dubrulle, L. Hartmann, N.  Calvet, D. Gautier, P. Michel, and P.
Tanga. R. Hueso acknowledges post-doctoral fellowships from
\textit{Gobierno Vasco} and the \textit{Observatoire de la
C\^{o}te d'Azur (bourse Poincar\'e)}. This work has been supported
by Spanish MCYT research project PNAYA2000-0932, the Universidad
del Pa\'{\i}s Vasco Grupos grant 13697/2001 and the French
\textit{Programme Nationale de Planetologie}. Early computations
were supported by the OCA program \textit{Simulations Interactives
et Visualisation en Astronomie et M\'ecanique (SIVAM)}. The final
calculations were done at the AJAX Cluster at the \textit{Grupo de
Ciencias Planetarias UPV/EHU} .
\end{acknowledgements}

\end{document}